\documentclass[floatfix,groupedaddress,superscriptaddress,aps,showpacs,amsmath,amssymb,floatfix, prx, longbibliography, twocolumn,a4]{revtex4-2}

\usepackage{graphicx,float}
\usepackage{subfigure}
\usepackage{dcolumn}
\usepackage{bm}
\usepackage{mathrsfs}
\usepackage{txfonts}
\usepackage{CJK}
\usepackage[amsmath]{SIunits}
\usepackage{epsfig}
\usepackage{epstopdf}
\usepackage{lipsum}
\usepackage{color}
\usepackage{placeins}

\usepackage[colorlinks,
linkcolor=blue,
anchorcolor=blue,
citecolor=blue,
filecolor=blue,
menucolor=blue,
runcolor=blue,
urlcolor=blue,
frenchlinks=blue]{hyperref}

\allowdisplaybreaks[2]
\newenvironment{SChinese}{%
\CJKfamily{gbsn}%
\CJKtilde
\CJKnospace}{}
\allowdisplaybreaks[2]

\begin{document}

\begin{CJK}{UTF8}{}
\begin{SChinese}

\title{Nonreciprocal single-photon band structure}

\author{Jiangshan Tang}  %
 \affiliation{College of Engineering and Applied Sciences, National Laboratory of Solid State Microstructures, and Collaborative Innovation Center of Advanced Microstructures, Nanjing University, Nanjing 210023, China}
  \affiliation{School of Physics, Nanjing University, Nanjing 210023, China}

  \author{Wei Nie}
 \affiliation{RIKEN Quantum Computing Center, RIKEN Cluster for Pioneering Research, Wako-shi, Saitama 351-0198, Japan}

\author{Lei Tang}  %
 \affiliation{College of Engineering and Applied Sciences, National Laboratory of Solid State Microstructures, and Collaborative Innovation Center of Advanced Microstructures, Nanjing University, Nanjing 210023, China}

 \author{Mingyuan Chen}  %
 \affiliation{College of Engineering and Applied Sciences, National Laboratory of Solid State Microstructures, and Collaborative Innovation Center of Advanced Microstructures, Nanjing University, Nanjing 210023, China}

  \author{Xin Su}  %
 \affiliation{College of Engineering and Applied Sciences, National Laboratory of Solid State Microstructures, and Collaborative Innovation Center of Advanced Microstructures, Nanjing University, Nanjing 210023, China}
  \affiliation{School of  Electronic Science and Engineering, Nanjing University, Nanjing 210023, China}

\author{Yanqing Lu}  %
     \affiliation{College of Engineering and Applied Sciences, National Laboratory of Solid State Microstructures, and Collaborative Innovation Center of Advanced Microstructures, Nanjing University, Nanjing 210023, China}
  \affiliation{School of Physics, Nanjing University, Nanjing 210023, China}

    \author{Franco Nori}
 \affiliation{RIKEN Quantum Computing Center, RIKEN Cluster for Pioneering Research, Wako-shi, Saitama 351-0198, Japan}
 \affiliation{Physics Department, The University of Michigan, Ann Arbor, Michigan 48109-1040, USA}

\author{Keyu Xia}  %
 \email{keyu.xia@nju.edu.cn}
    \affiliation{College of Engineering and Applied Sciences, National Laboratory of Solid State Microstructures, and Collaborative Innovation Center of Advanced Microstructures, Nanjing University, Nanjing 210023, China}
  \affiliation{School of Physics, Nanjing University, Nanjing 210023, China}
  \affiliation{Jiangsu Key Laboratory of Artificial Functional Materials, Nanjing University, Nanjing 210023, China}

\date{\today}

\begin{abstract}
We study single-photon band structure in a one-dimensional (1D) coupled-resonator optical waveguide (CROW) which chirally couples to an array of two-level quantum emitters (QEs). The chiral interaction between the resonator mode and the QE can break the time-reversal symmetry without the magneto-optical effect. As a result, a nonreciprocal single-photon edge state, band gap and flat band appear. By using such a chiral QE-CROW system, including a finite number of unit cells and working in the nonreciprocal band gap, we achieve frequency-multiplex single-photon circulators with high fidelity and low insertion loss. The chiral QE-light interaction can also protect one-way propagation of single photons against backscattering. Our work opens a new door for studying nonreciprocal photonic band structure and exploring its applications in the quantum regime.
\end{abstract}

\maketitle

\end{SChinese}
\end{CJK}

\emph{Introduction}.---Optical nonreciprocity plays an indispensable role in many important applications including quantum-information processing \cite{nature07127}, invisible sensing \cite{PhysRevLett.106.213901,natmater.18.8}, noise-free information processing \cite{nphoton.2013.185} and unconventional lasing \cite{science.aao4551,PhysRevApplied.10.064037}.
Conventional nonreciprocal optical devices breaking the time-reversal symmetry rely on the weak magneto-optical effect and are difficult to be integrated on a chip due to the constraint of a strong magnetic field.

Various theoretical schemes and experimental methods have been reported to circumvent the constraint of magnetic fields by using optical nonlinearity \cite{nphysics.10.5,  PhysRevLett.118.033901, nphoton.2018.1038, PhysRevLett.121.203602, PhysRevLett.123.233604,PhysRevResearch.2.033517,PRJ.413286,tanglei.3970886}, optomechanical resonators \cite{OE.20.007672,nphoton.2016.161,ncomms13662}, spinning resonators \cite{natures41586.018,PhysRevLett.121.153601}, tempo-spatial modulation of the medium \cite{PhysRevLett.110.093901,PhysRevLett.110.223602,nphotons41566.017} or chiral quantum optics systems \cite{PhysRevX.5.041036,PhysRevA.90.043802,science.aaj2118, nature21037,OPTICA.393035, sciadv.abe8924,ncomun.10.1038.2389,pucher2021atomic, PhysRevA.99.043833}.
In contrast to classical methods, a chiral quantum optics system
can realize optical nonreciprocity in the quantum regime \cite{PhysRevA.90.043802, science.aaj2118, nature21037, PhysRevLett.121.203602, nphoton.2018.1038, PhysRevLett.123.233604,sciadv.abe8924} and can perform chiral quantum information processing \cite{PhysRevLett.113.237203, nnano.2015.159,OE.29.11.17613,PhysRevApplied.15.064020}.

It is of great interest to study photon behavior under the chiral interaction between quantum emitters (QEs) and a coupled resonator array. The chiral interaction of some QEs with few resonators has been extensively studied and promises many important applications ~\cite{RevModPhys.89.021001,RevModPhys.90.031002}. Although the coupled-resonator optical waveguide (CROW) has been widely investigated \cite{OL.24.000711,OE.12.000090,PhysRevLett.124.083603, PhysRevLett.125.013902, APL.119.14.141108}, the optical chirality in the CROW is barely addressed. In analogy to the electronic band structure in condensed matter, a periodic photonic structure exhibits nontrivial band structures and allows defect-immune photon transport \cite{PhysRevB.84.075477,nphoton.2014.248,nphotons.11.12,RevModPhys.91.015006,sciadv.abe1398}. Nevertheless,  such topological optical system can break time-reversal symmetry and is appealing for backscattering-immune optical isolation only in the presence of an external/synthetic magnetic field. Thus, it is highly desirable to achieve one-way photonic band structure such as nonreciprocal single-photon edge states (SPESs) and band gap without magnetic fields, in particular, in the \emph{quantum} regime.

By exploring the chiral interaction between an array of QEs and a 1D CROW, here we show that \emph{nonreciprocal single-photon band structure, including SPES, band gap and flat band, can be achieved without a magnetic field.}
Interestingly, we can conduct \emph{frequency-multiplex single-photon circulators} with such a chiral QE-CROW system working in the nonreciprocal band gap. Our circulator is \emph{immune to backscattering}.

\begin{figure}
  \centering
  \includegraphics[width=1.0\linewidth]{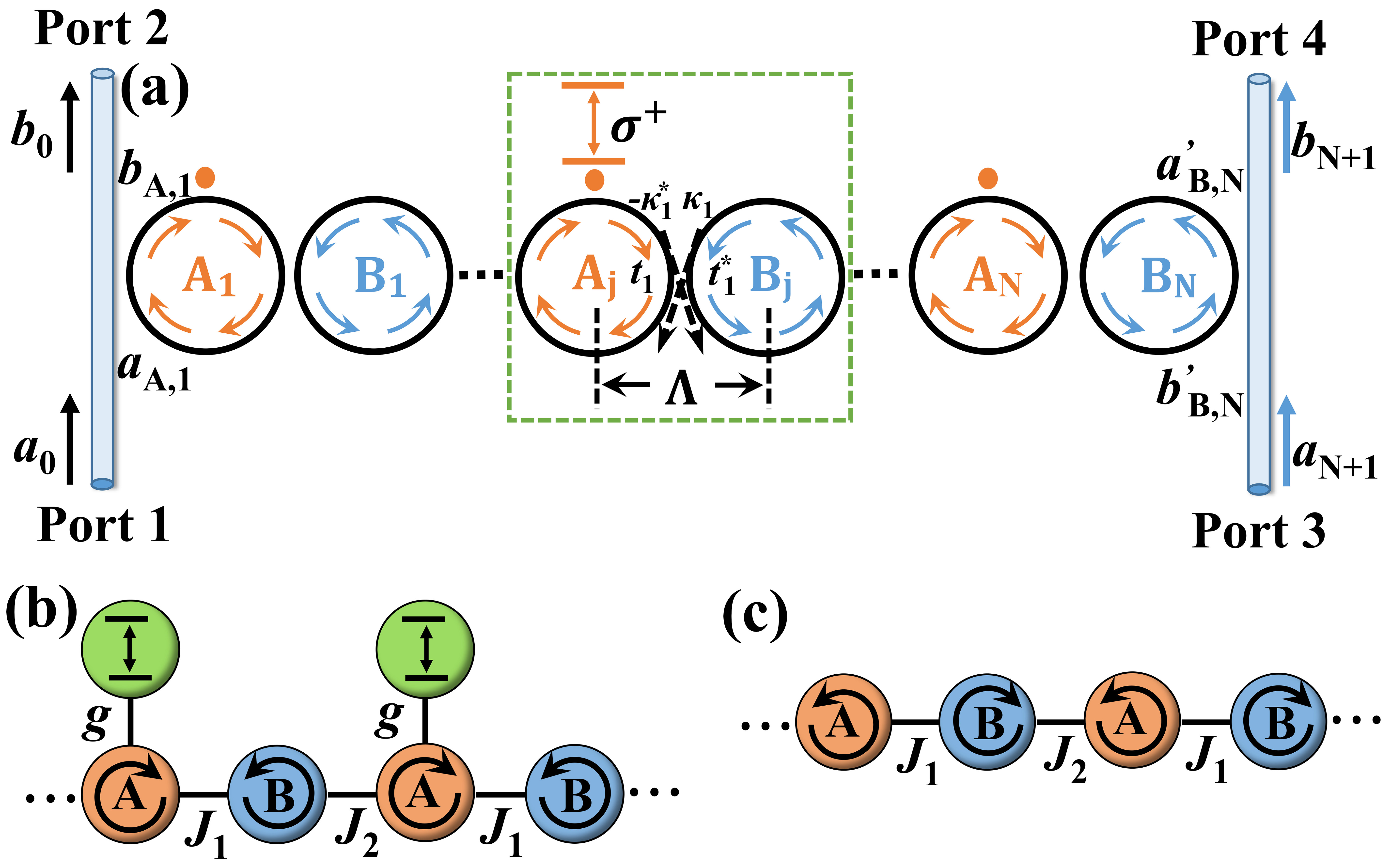} \\
\caption{(a) Schematic of a chiral QE-CROW system containing $N$ unit cells and the input and output waveguides. The arrows represent the circulating direction of the whispering-gallery mode for an input to port 1 or 3 (forward input). The photon incident to ports 2 and 4 (backward input) excites the oppositely propagating resonator modes. The QEs with $\sigma^{+}$-polarized optical transition are periodically coupled to the clockwise resonator mode of the sublattice A in each unit cell, see dashed box. (b) The L-type trimer chain in the forward-input case. (c) The dimer chain for the backward-input case.}
\label{fig:FIG1}
\end{figure}
%
\emph{System and model}.---The schematic of the chiral QE-CROW system is depicted in Fig.~\ref{fig:FIG1}. In the CROW, the evanescent field of each microring resonator is almost perfectly circularly polarized and thus possesses nearly unitary optical chirality \cite{PhysRevA.90.043802,science.aaj2118,PhysRevA.99.043833}, i.e. possessing spin-momentum locking \cite{PhysRevA.83.021803,ncomms.5.1,science.348.1148,PhysRevLett.120.243605}. In each unit cell, the adjacent resonators are separated by $\Lambda$ and coherently coupled via evanescent fields. Each resonator supports two degenerate optical whispering-gallery modes circulating along either the clockwise (CW) or counterclockwise (CCW) direction. We divide the resonators into A and B sublattice groups, respectively denoted as $\text{R}_\text{A}$ and $\text{R}_\text{B}$. In each unit cell, a two-level QE with frequency $\omega_q$ is only embedded in the A-sublattice resonator $\text{R}_\text{A}$, but decouples from the B-sublattice resonator $\text{R}_\text{B}$. Within a cell, the CW (CCW) mode of $\text{R}_\text{A}$ couples to  the CCW (CW) mode of $\text{R}_\text{B}$ with strength $J_1$, see the green box in Fig.~\ref{fig:FIG1}(a). The $j$th resonator $\text{R}_\text{A}$ couples to the $(j+1)$th resonator $\text{R}_\text{B}$ in the next cell with strength $J_2$. The oppositely circulating whispering-gallery modes in the sublattices A and B form $N$ unit cells with a lattice constant of $2\Lambda$ \cite{OE.12.000090,JOSAB.21.001665}. We assume that the resonators, with resonance frequency $\Omega$ and internal dissipation rate $\gamma_{\text{in}}$, are identical.

Hereafter, we consider only a single excitation in the QE-CROW system. The QEs can be precisely positioned atoms \cite{science.aaj2118,science.1237125,PhysRevLett.124.083603}, quantum dots \cite{RevModPhys.90.031002,science.aaq0327,s41377-020-0244-9}, or nanopillars covered by monolayers \cite{ncomms15053,ncomms15093}. By optically initializing the QEs in specific spin ground states or shifting the transition energy with
a circularly-polarized field via the optical Stark effect, we can induce a chiral transition in the QE. Without loss of generality, we assume that the QE transition and the evanescent field of the CW mode of $\text{R}_\text{A}$ are both $\sigma_+ -$polarized. In this arrangement, we create a chiral interaction between the QE and the CW mode of $\text{R}_\text{A}$ without using a magnetic field \cite{PhysRevA.90.043802,PhysRevX.5.041036,science.aaj2118,nature21037,ncomun.10.1038.2389,PhysRevA.99.043833,pucher2021atomic}. However, the QE decouples from the CCW mode~\cite{PhysRevA.90.043802,science.aaj2118,nature21037,PhysRevA.99.043833}.

When a single photon is incident into port 1 or 3, namely the forward-input case, the CW modes in the $\text{R}_\text{A}$ resonators and the CCW ones in the $\text{R}_\text{B}$ resonators are driven. We refer to this excitation as the $\text{CW}_A-\text{CCW}_B$ supermode ( opposite to the $\text{CCW}_A-\text{CW}_B$ supermode). In this case, the QEs are coupled to $\text{R}_\text{A}$ in each unit cell. The QE-CROW system is equivalent to an L-type trimer chain with each cell containing three sublattices, see Fig.~\ref{fig:FIG1}(b). We take $\omega_q=\Omega$ and can write the Hamiltonian of the system in the rotating frame as:
\begin{equation}\label{eq:HamiCW}
  \begin{split}
    \hat{ H}_{\text{CW}_A-\text{CCW}_B} &=  \sum_{j}^{N}\left(g\hat{a}^{\dag}_{j,\circlearrowright}\hat{\sigma}_{j} + J_1 \hat{a}^{\dag}_{j,\circlearrowright}\hat{b}_{j,\circlearrowleft} + \text{H.c.}\right) \; \\
       & + \sum_{j}^{N-1}\left( J_2 \hat{a}^{\dag}_{j+1,\circlearrowright}\hat{b}_{j,\circlearrowleft} + \text{H.c.}\right) \;,
  \end{split}
\end{equation}
where $\hat{a}_{j,\circlearrowright}$ ($\hat{b}_{j,\circlearrowleft}$) is the annihilation operator of the CW (CCW) mode of the $\text{R}_\text{A}$ ($\text{R}_\text{B}$) resonator in the $j$th unit cell, and $\hat{\sigma}_j$ denotes the lowering operator of the $j$th two-level QE.

Now we study the band structure of an infinite 1D-CROW model in a single-excitation subspace. We assume a periodic boundary condition along the 1D-CROW chain and apply the Fourier transform, $\zeta_{j}=\left(1/\sqrt{N}\sum_{k}e^{ikj}\zeta_{k}\right)$ with $\zeta_{j}=\left(\hat{a}_{j,\circlearrowright},\hat{\sigma}_{j}, \hat{b}_{j,\circlearrowleft}\right)^{T}$, to Eq.~\eqref{eq:HamiCW}, thus, $ \hat{ H}_{\text{CW}_A-\text{CCW}_B}=\sum_{k}\zeta_{k}^{\dag}\hat{ H}_{\text{CW}_A-\text{CCW}_B}(k)\zeta_{k}$. Then, the Hamiltonian takes the following form in wavevector space
\begin{equation}\label{eq:HamiCWk}
\hat{H}_{\text{CW}_A-\text{CCW}_B}\left(k\right)=\left(\begin{array}{ccc}
0 & g & J_{1}+J_{2} e^{-i k} \\
g & 0 & 0 \\
J_{1}+J_{2} e^{i k} & 0 & 0
\end{array}\right)\;.
\end{equation}

In comparison, if a single photon incidents to port 2 or 4 corresponding to the backward-input case, then the $\text{CCW}_A-\text{CW}_B$ supermode is driven in each unit cell and the QEs decouple from all resonators. Our system reduces to a dimer chain, corresponding to a standard Su-Schrieffer-Heeger (SSH) model \cite{PhysRevLett.42.1698,PhysRevLett.127.147401}, see Fig.~\ref{fig:FIG1}(c). The Hamiltonian then becomes
\begin{equation}\label{eq:HamiCCWk}
\hat{H}_{\text{CCW}_A-\text{CW}_B}\left(k\right)=\left(\begin{array}{cc}
0  & J_{1}+J_{2} e^{-i k} \\
J_{1}+J_{2} e^{i k} & 0
\end{array}\right)\;.
\end{equation}

The band structures and energy spectra of the system in the $\text{CW}_A-\text{CCW}_B$ and $\text{CCW}_A-\text{CW}_B$ supermodes can be found by solving the eigenvalues and eigenstates of Eq.~(\ref{eq:HamiCWk}) and Eq.~(\ref{eq:HamiCCWk}), respectively. In the case related to the $\text{CW}_A-\text{CCW}_B$ supermode, the QEs interact with the $\text{R}_\text{A}$ resonators. This interaction causes a phase and amplitude modulation $e^{i\varphi}$ to the $\text{R}_\text{A}$ $\text{CW}$ mode~\cite{SupplMat,tang2021}. Here, $\varphi$ is a complex number. In contrast, the QE-resonator interaction is absent for the $\text{CCW}_A-\text{CW}_B$ supermode. Thus, the extra modulation disappears, i.e. $\varphi=0$. The dispersion relations of the $\text{CW}_A-\text{CCW}_B$ supermode consists of three dispersive bands. It is essentially different from the $\text{CCW}_A-\text{CW}_B$ supermode, which only contains two bands. Therefore, the single-photon band structures are nonreciprocal in two counter-propagating cases. This nonreciprocal single-photon dispersion occurs because the chiral QE-light coupling breaks the time-reversal symmetry of the system.

A finite 1D-CROW chain with  a large number of unit cells can approximately exhibit the property of an infinite chain \cite{OE.12.000090}. Next, we discuss a finite chiral 1D CROW containing $N$ unit cells ($N\geq1$) and evaluate the single-photon transmission. Two optical waveguides are side-coupled to the terminal resonators as input and output channels. We utilize the transfer matrix method to investigate the transmission properties of the system~\cite{PhysRevLett.125.013902,SupplMat}. We consider the forward case. The notations for the field components $\{a\}$ and $\{b\}$ are shown in Fig.~\ref{fig:FIG1}(a). By successively multiplying the transfer matrices of all the coupling relations, we obtain the transport relation
\begin{equation}
  \label{eq:input}
 \left(\begin{array}{l}
       a_{N+1} \\
       b_{N+1}
  \end{array}\right) =M\left(\begin{array}{l}
       a_{0} \\
       b_{0}
  \end{array}\right) \equiv \left(\begin{array}{cc}
     M_{11}& M_{12} \\
     M_{21}& M_{22}
  \end{array}\right)\left(\begin{array}{c}
       a_{0} \\
       b_{0}
  \end{array}\right)  \;,
  \end{equation}
$M = M_{\text{out}} M_{\text{p,B}} M_{\text{c,1}} M_{\text{p,A}} (M_{\text{c,2}} M_{\text{p,B}} M_{\text{c,1}} M_{\text{p,A}})^{N-1}M_{\text{in}}$, where
\begin{equation}
\begin{aligned}
&M_{\text{p,A}}=\left(\begin{array}{cc}
\alpha e^{-i \theta} & 0 \\
0 & \alpha e^{i(\theta+\varphi)}
\end{array}\right), \quad M_{\text{p,B}}=\left(\begin{array}{cc}
\alpha e^{-i \theta} & 0 \\
0 & \alpha e^{i \theta}
\end{array}\right), \\
&M_{\text{c,1}}=\frac{1}{\kappa_{1}}\left(\begin{array}{cc}
1 & -t_{1} \\
t_{1}^{*} & -1
\end{array}\right), \quad M_{\text{c, 2}}=\frac{1}{\kappa_{2}}\left(\begin{array}{cc}
1 & -t_{2} \\
t_{2}^{*} & -1
\end{array}\right), \\
&M_{\text{in}}=\frac{1}{\kappa_{\text{in}}}\left(\begin{array}{cc}
-t_{\text{in}} & 1 \\
-1 & t_{\text{in}}^{*}
\end{array}\right), \quad M_{\text{out}}=\frac{1}{\kappa_{\text{out}}}\left(\begin{array}{cc}
1 & -t_{\text{out}} \\
t_{\text {out }}^{*} & -1
\end{array}\right) .
\end{aligned}
\end{equation}
We define $\{t_{\text{in}},t_{1},t_{2},t_{\text{out}}\}$ and $\{\kappa_{\text{in}},\kappa_1,\kappa_2,\kappa_{\text{out}}\}$ as the transmission and coupling coefficients of the intrasite and intersite resonators, where $t_i~(i=\text{in},1, 2, \text{out})$ is a real number, $\kappa_i$ is an imaginary number, and $|t_i|^2+|\kappa_i|^2=1$ \cite{PhysRevLett.125.013902}. Above, $\theta$ indicates the phase change and $\alpha$ is the loss coefficient of propagating single photons. This can be calculated as $\alpha \approx 1-2\gamma_{\text{in}}/\mathcal{F}$ from the intrinsic loss rate $\gamma_{\text{in}}$ of the resonator \cite{tang2021}, where $\mathcal{F}$ is the free spectral range of the resonator considered here. The ratio $\kappa_{\text{in}}/\mathcal{F}$ is negligible for our high-quality resonators. Thus, hereafter we can take $\alpha\approx1$.

For simplicity, we assume $\kappa_{\text{in}}=\kappa_{\text{out}}$ ($t_{\text{in}} = t_\text{out}$). The external losses of the edge resonators $A_1$ and $B_\text{N}$ are $\gamma_{\text{ex}}=\gamma_{\text{ex,1}}=\gamma_{\text{ex,N}}=-\text{ln}\left(|t_{\text{in}}|\right)\times\mathcal{F}$. Considering a single-photon entering port 1 ($a_{N+1}=0$), we obtain the transmission matrix elements between the input and output ports:
\begin{subequations}
  \label{eq:TR1}
  \begin{align}
    T_{12} & = \left|\frac{b_0}{a_0}\right|^2=\left|\frac{M_{11}}{M_{12}}\right|^2 \;, \\
    T_{14} & = \left|\frac{b_{N+1}}{a_0}\right|^2=\left|M_{21} - \frac{M_{11}M_{22}}{M_{12}}\right|^2 \;,
  \end{align}
 \end{subequations}
where $T_{mn}$ is the transmission from port $m$ to port $n$, with $m,n=1,2,3,4$. Because the input to port 3 excites the $\text{CW}_A-\text{CCW}_B$ supermode circulating along the same direction, we have $T_{12}=T_{34}$ and $T_{14}=T_{32}$. When the $\text{CCW}_A-\text{CW}_B$ supermode is excited in the backward case, the QEs decouple from the resonators and we have $\varphi=0$. Therefore, we have $T_{mn} \neq T_{nm}$ for $m\neq n$, indicating the occurrence of optical nonreciprocity in our chiral QE-CROW system.

Below, we consider an ideal scenario in which the QE dissipation rate $\gamma$ is negligible compared to the coupling strength $g$. For simplicity, we set $\gamma = 0$.

\begin{figure}
  \centering
  \includegraphics[width=1.0\linewidth]{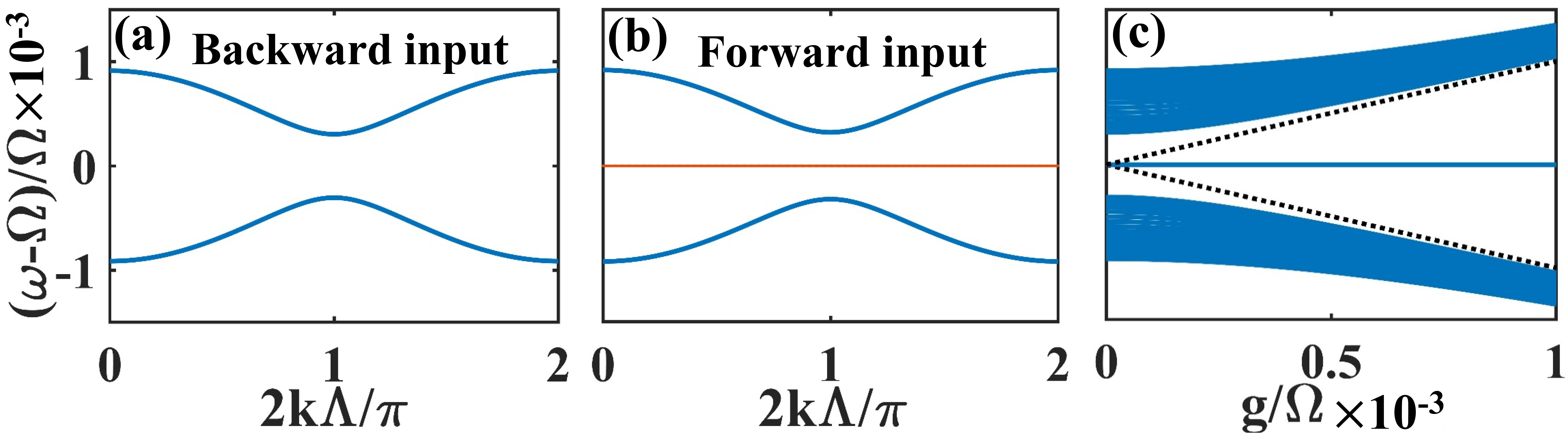} \\
  \caption{Single-photon band structures of the chiral QE-CROW system in the case of $J_{1}<J_{2}$. (a) Band structure when the $\text{CCW}_A-\text{CW}_B$ supermode is excited in the absence of the QEs, i.e., $g/\Omega=0$. (b) Band structure when the $\text{CW}_A-\text{CCW}_B$ supermode, coupling to the QEs with $g/\Omega=10^{-4}$, is driven.  (c) Band structure versus the coupling strength $g$ when the $\text{CW}_A-\text{CCW}_B$ supermode is driven. Other parameters are $J_{1}/\Omega=3\times10^{-4}$, $J_{2}=2J_{1}$.}
\label{fig:FIG2}
\end{figure}

\begin{figure}
  \centering
  \includegraphics[width=1.0\linewidth]{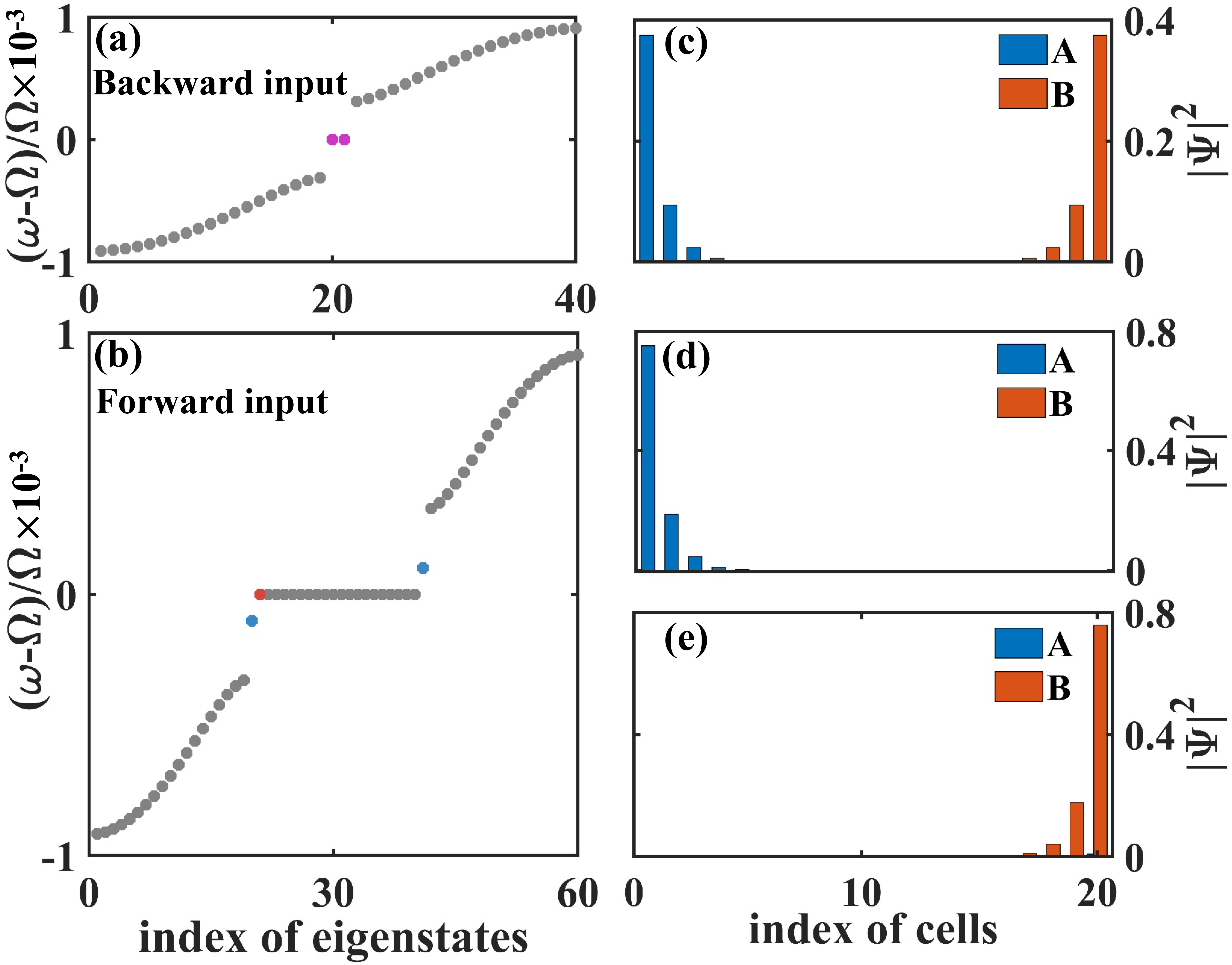} \\
  \caption{Nonreciprocal energy spectra (a, b) and probability distribution (c-e). (a) and (c) for the $\text{CCW}_A-\text{CW}_B$ supermode without coupling to the QEs, $g=0$. (b), (d), and (e) for the $\text{CW}_A-\text{CCW}_B$ supermode with coupling to the QEs, $g/\Omega=10^{-4}$. Other parameters are $J_{1}/\Omega=3\times10^{-4}$, $J_{2}=2J_{1}$.}
\label{fig:FIG3}
\end{figure}

\emph{Nonreciprocal single-photon flat band and edge states}.---We first discuss the case $J_1<J_2$.  As seen from the dispersion relations in Fig.~\ref{fig:FIG2},  the QE-CROW system exhibits a nonreciprocal single-photon band structure. Typically, two dispersive bands of the SSH model appear in the backward case, as shown in Fig.~\ref{fig:FIG2}(a) \cite{asboth2016short} for $J_{1}/\Omega=3\times10^{-4}$ and $J_{2}=2J_{1}$. The curves with opposite slopes imply two opposite directions of single-photon propagation, corresponding to the input to port 2 or 4 in Fig.~\ref{fig:FIG1}, respectively. In the forward case, the QEs interact with the $\text{R}_\text{A}$ resonator. A flat band appears at $\omega=\Omega$ due to the coupling to the QEs, dividing the original band gap into two parts, see Fig.~\ref{fig:FIG2}(b). Thus, our QE-CROW system allows \emph{slow light and delay of single-photon pulses} \cite{PhysRevLett.93.233903,nature.577.7788,PhysRevLett.126.103601}.
 As the QE-resonator coupling strength increases, the two band gaps become wider, see Fig.~\ref{fig:FIG2}(c). Interestingly, these two band gaps possess very peculiar SPESs. These SPESs are very different from the backward case.

 Figure~\ref{fig:FIG3} shows energy spectra of a finite unit cell with $N=20$ and the probability distribution corresponding to the SPESs in the two oppositely circulating supermodes of the system. As for the $\text{CCW}_A-\text{CW}_B$ supermode without coupling to the QEs, there are $2N$ eigenvalues and two degenerate zero-energy SPESs falling in the bulk band gap [see the purple asterisks in Fig.~\ref{fig:FIG3}(a)]. The wave functions of two edge states localize at the left or right boundary and exponentially decay from the boundaries [see Fig.~\ref{fig:FIG3}(c)].

 In stark contrast, when the system is excited in the $\text{CW}_A-\text{CCW}_B$ supermode, the coupling to the QEs results in $N$ more eigenvalues. Due to the coupling with the QEs, the left SPES becomes doublet, forming two superstates of the $\text{R}_A$ set and the QEs with an energy splitting proportional to the coupling strength $g$ \cite{PhysRevB.95.235143,PhysRevA.99.013833,SupplMat}, see the dashed black curves in Fig.~\ref{fig:FIG2}(c). As an example, the energy spectrum of the QE-CROW system with $N=20$ and  $g/\Omega=10^{-4}$ is shown in Fig.~\ref{fig:FIG3}(c). The left SPESs have eigenenergies at $\omega-\Omega=\pm g$ (see the blue asterisks). In Fig.~\ref{fig:FIG3}(d), the probability of the A sublattice includes the contribution of the QE. Unlike the SPESs of the $\text{CCW}_A-\text{CW}_B$ supermode, the wave functions of these SPESs only localize on the left edge of the system with an exponentially decaying probability distribution at the sites of both the A-sublattice resonator and the QE. The right zero-energy SPES still exists, [see the red asterisk in Fig.~\ref{fig:FIG3}(b)], because the $\text{R}_\text{B}$ resonators decouple from the QE. The wave function of this right SPES localizes at the right boundary [see Fig.~\ref{fig:FIG3}(e)]. The nonreciprocal SPESs is the unique property of our chiral QE-CROW system.

For small $N$, the left and right SPESs overlap in the backward-input case because of their exponentially decaying distribution. This wavefunction overlap at zero-energy frequency results in a SPES tunneling. In contrast, in the forward case, this SPES-overlap induced tunneling is greatly suppressed because the left and right SPESs are nondegenerate in energy. Therefore, we can realize a single-photon circulator via the nonreciprocal SPESs with few unit cells \cite{SupplMat}.

\begin{figure}
  \centering
  \includegraphics[width=1.0\linewidth]{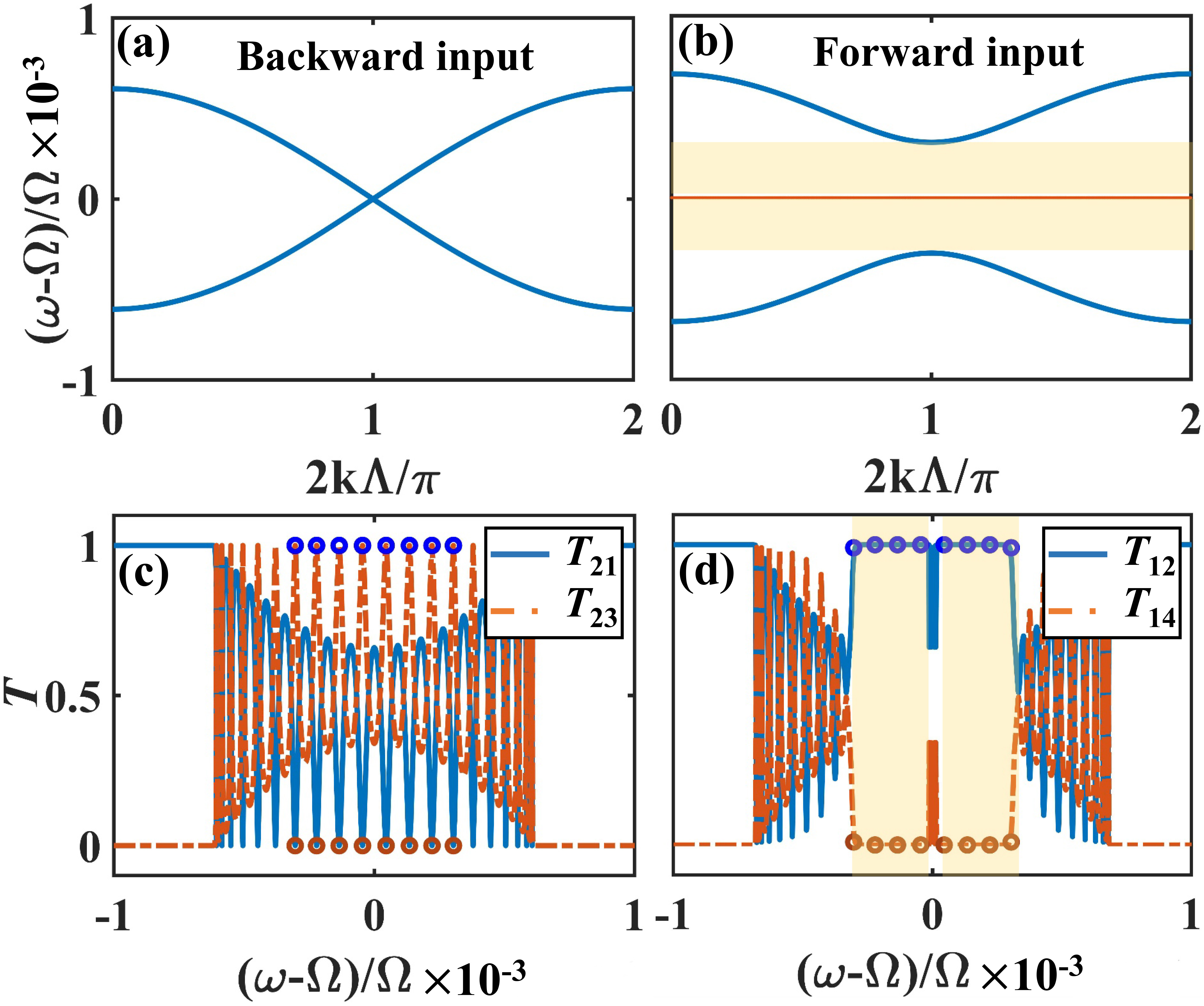} \\
  \caption{Single-photon band structure (a, b) and transmission (c, d) with $N=10$ for $J_1=J_2$. In (a, c) the $\text{CCW}_A-\text{CW}_B$ supermode is excited, and $g =0$. In (b, d) the $\text{CW}_A-\text{CCW}_B$ supermode is driven, and $g/\Omega =3\times10^{-4}$. Other parameters are $J_{1}/\Omega=3\times10^{-4}$, $J_{2}=J_{1}$.}
\label{fig:FIG4}
\end{figure}
\emph{Nonreciprocal single-photon band gap}.---The case $J_{1}>J_{2}$ opens a band gap but exhibits trivial physical properties. We are not interested in it here. Now we discuss the case $J_{1}=J_{2}$. Figures~\ref{fig:FIG4}(a,b) show a nonreciprocal single-photon band gap for oppositely propagating photons. For the backward input, the two bands are closed and thus form continuous conduction bands. In comparison, as in the case of $J_{1} < J_{2}$, the $\text{CW}_A-\text{CCW}_B$ supermode also exhibits two band gaps, separated by a flat band due to the interaction with QEs, see the yellow region in Fig.~\ref{fig:FIG4}(b). The band gap increases with the QE-resonator coupling strength, permitting \emph{a wider nonreciprocal bandwidth}. This nonreciprocal single-photon band gap allows to conduct \emph{single-photon circulators}.
The same results can be derived by using the transfer matrix method \cite{SupplMat}.

The band structures of our system characterize the transmission of single photons. When we consider the case with finite unit cells in the 1D CROW, the conduction band region turns into transmission peaks. Using Eqs.~(\ref{eq:TR1}), we calculate the transmission spectrum of our QE-CROW system with $N=10$ in the two opposite-propagation cases.
For single photons input to port 2, see Fig.~\ref{fig:FIG4}(c), the excited $\text{CCW}_\text{A} - \text{CW}_\text{B}$ supermode decouples from the QEs. This system is equivalent to the conventional CROW without QEs. There are $2N$ peaks in the transmission $T_{23}$ and dips in $T_{21}$.
In contrast, when the single-photon wave packet inputs to port 1, the transmission $T_{14}$ is zero, because the photon transport is forbidden in the band gap, and thus $T_{12}=1$, see Fig.~\ref{fig:FIG4}(d). Clearly, we obtain \emph{strong nonreciprocity in the transmission} at the frequencies corresponding to the peaks of $T_{23}$, where $T_{23} \approx 1$, or the dips of $T_{21}$, where $T_{21} \approx 0$. For $g/\Omega =3\times 10^{-4}$, we have eight nonreciprocal frequency windows at $\left(\omega-\Omega\right)/\Omega \approx \{\pm30,\pm22,\pm13,\pm4.5\} \times 10^{-5}$, see blue circles.
Due to the symmetry of the CROW, we obtain $T_{12}=T_{34}$ and $T_{23}=T_{41}$. Thus, we can choose any one of eight frequencies in the band-gap region to achieve $T_{23}=1$ ($T_{21}=0$). In this case, we can attain \emph{a frequency-multiplex single-photon circulator} with a circling photon transport direction $1\rightarrow2\rightarrow3\rightarrow4\rightarrow1$ at frequencies marked by circles in Figs.~\ref{fig:FIG4}(d).

Note that this single-photon circulator is robust against backscattering (see the Supplemental Material \cite{SupplMat}). The chiral coupling between the QE and the vacuum field of  $\text{R}_\text{A}$ eigenmodes induces an extra phase shift $\text{Re}[\varphi]$ for the CW mode, shifting its resonance frequency and leading to non-degenerate with the CCW mode.
Thus, the opposite CW mode excited by a backscattering of a propagating CCW mode is forbidden in the nonreciprocal band gap, see Figs.~\ref{fig:FIG4} (a,b). Our chiral QE-CROW system promises a new type of backscattering-immune optical device.

\emph{Implementation.}---The required 1D CROW can be made with silicon oxynitride \cite{OL.33.002389}, silicon on insulator \cite{nphoton.2013.274,fmats.2015.00034} or lithium niobium oxynitride \cite{NJP.22.7.2020}. We consider a QE-CROW system consisting of $N=10$ resonators with a radius of $r = 40~\micro\meter$ and $n_{\text{eff}}=2$, yielding $\mathcal{F}/2\pi=0.6~\tera\hertz$. The external dissipations of resonators can be calculated as $\gamma_{\text{ex}}/2\pi=19.4~\giga\hertz$ [setting $\gamma_{\text{in}}=0.02\gamma_{\text{ex}}$, thus $\gamma_{\text{tol}}/2\pi=(\gamma_{\text{ex}}+\gamma_{\text{in}})/2\pi=19.8~\giga\hertz$] \cite{tang2021}. For the coupling coefficient $\kappa_1=\kappa_2=0.1i$, the coupling strength between resonators can be deduced as $J_{1}/\gamma_{\text{tol}}=J_{2}/\gamma_{\text{tol}}=3$. We take $\Omega/2\pi=195~\tera\hertz$. Regarding QEs, we choose $\lambda=1.538~\micro\meter$ \cite{apl.112.17.2018,nanolett.18.5.2018,APR.7.2.2020} so that $\omega_q=\Omega$ and a dipole moment of $|\textbf{d}|=20~\text{Debye}$ \cite{PhysRevLett.88.087401}, yielding $\gamma=2\pi \times 5.48~\mega\hertz$, and $g/\gamma_{\text{tol}}=0.77$ for a resonator mode volume of $V_{\text{m}}=2.0~\micro\meter^3$ \cite{Xu:08,PhysRevA.99.043833}. The chiral QE-light interaction can be optically induced as aforementioned.
The fidelity and the average photon survival probability of the circulator are calculated as $\{0.99,0.98\}$ at two optimal frequency windows $(\omega-\Omega)/\Omega \approx \pm4.5\times10^{-5}$ \cite{science.aaj2118,PhysRevLett.121.203602}. Moreover, we achieve a nonreciprocal transmission bandwidth of $\sim 2\pi\times 2.4~\giga\hertz$ and an average insertion loss of $~1.12~\deci\bel$ for a circulator fidelity larger than $0.95$. Thus, the nonreciprocal SPES and high-performance single-photon circulator can be achieved even in the weak coupling regime $g < \gamma_\text{tol}$. In stark contrast, a conventional chiral quantum optics system using a single resonator can achieve the single-photon diode and circulator only in the strong coupling regime~\cite{PhysRevA.90.043802,science.aaj2118,nature21037,PhysRevA.99.043833}. Thus, our proposal greatly relaxes its experimental challenge.

\emph{Conclusion.}---We have shown that nonreciprocal single-photon band structures can be achieved in a chiral QE-CROW system. The chiral QE-CROW system with a nonreciprocal band gap can construct frequency-multiplex single-photon circulators. Our work extends photonic band structures of periodic photonic structures to exhibit remarkable magnetic-free quantum nonreciprocity, likely beyond condensed matter.

K.X. and Y.L thank Prof. Haiqing Lin for his stimulating and constructive discussion. J.T. thanks Chaohua Wu for helpful discussions. Y.L. and K.X. are supported by the National Key R\&D Program of China (Grants No. 2019YFA0308700, No. 2017YFA0303703, and No. 2017YFA0303701), the National Natural Science Foundation of China (Grants No. 11874212 and No. 11890704), the Fundamental Research Funds for the Central
Universities (Grant No. 021314380095), the Program for Innovative Talents and Entrepreneurs in Jiangsu, and the Excellent Research Program of Nanjing University (Grant No. ZYJH002). W.N is supported by the National Natural Science Foundation of China (Grants No. 11690031).  F.N. is supported by the Nippon Telegraph and Telephone Corporation (NTT) Research, the Japan Science and Technology Agency (JST) [via the Quantum Leap Flagship Program (Q-LEAP), the Moonshot R\&D Grant Number JPMJMS2061, and the Centers of Research Excellence in Science and Technology (CREST) Grant No. JPMJCR1676], the Japan Society for the Promotion of Science (JSPS) [via the Grants-in-Aid for Scientific Research (KAKENHI) Grant No. JP20H00134 and the JSPS-RFBR Grant No. JPJSBP120194828], the Army Research Office (ARO) (Grant No. W911NF-18-1-0358), the Asian Office of Aerospace Research and Development (AOARD) (via Grant No. FA2386-20-1-4069), and the Foundational Questions Institute Fund (FQXi) via Grant No. FQXi-IAF19-06. We thank the High Performance Computing Center of Nanjing University for allowing the numerical calculations on its blade cluster system.


\begin{thebibliography}{76}%
\makeatletter
\providecommand \@ifxundefined [1]{%
 \@ifx{#1\undefined}
}%
\providecommand \@ifnum [1]{%
 \ifnum #1\expandafter \@firstoftwo
 \else \expandafter \@secondoftwo
 \fi
}%
\providecommand \@ifx [1]{%
 \ifx #1\expandafter \@firstoftwo
 \else \expandafter \@secondoftwo
 \fi
}%
\providecommand \natexlab [1]{#1}%
\providecommand \enquote  [1]{``#1''}%
\providecommand \bibnamefont  [1]{#1}%
\providecommand \bibfnamefont [1]{#1}%
\providecommand \citenamefont [1]{#1}%
\providecommand \href@noop [0]{\@secondoftwo}%
\providecommand \href [0]{\begingroup \@sanitize@url \@href}%
\providecommand \@href[1]{\@@startlink{#1}\@@href}%
\providecommand \@@href[1]{\endgroup#1\@@endlink}%
\providecommand \@sanitize@url [0]{\catcode `\\12\catcode `\$12\catcode
  `\&12\catcode `\#12\catcode `\^12\catcode `\_12\catcode `\%12\relax}%
\providecommand \@@startlink[1]{}%
\providecommand \@@endlink[0]{}%
\providecommand \url  [0]{\begingroup\@sanitize@url \@url }%
\providecommand \@url [1]{\endgroup\@href {#1}{\urlprefix }}%
\providecommand \urlprefix  [0]{URL }%
\providecommand \Eprint [0]{\href }%
\providecommand \doibase [0]{https://doi.org/}%
\providecommand \selectlanguage [0]{\@gobble}%
\providecommand \bibinfo  [0]{\@secondoftwo}%
\providecommand \bibfield  [0]{\@secondoftwo}%
\providecommand \translation [1]{[#1]}%
\providecommand \BibitemOpen [0]{}%
\providecommand \bibitemStop [0]{}%
\providecommand \bibitemNoStop [0]{.\EOS\space}%
\providecommand \EOS [0]{\spacefactor3000\relax}%
\providecommand \BibitemShut  [1]{\csname bibitem#1\endcsname}%
\let\auto@bib@innerbib\@empty
\bibitem [{\citenamefont {Kimble}(2008)}]{nature07127}%
  \BibitemOpen
  \bibfield  {author} {\bibinfo {author} {\bibfnamefont {H.~J.}\ \bibnamefont
  {Kimble}},\ }\bibfield  {title} {\bibinfo {title} {The quantum internet},\
  }\href {https://doi.org/10.1038/nature07127} {\bibfield  {journal} {\bibinfo
  {journal} {Nature (London)}\ }\textbf {\bibinfo {volume} {453}},\ \bibinfo
  {pages} {1023} (\bibinfo {year} {2008})}\BibitemShut {NoStop}%
\bibitem [{\citenamefont {Lin}\ \emph {et~al.}(2011)\citenamefont {Lin},
  \citenamefont {Ramezani}, \citenamefont {Eichelkraut}, \citenamefont
  {Kottos}, \citenamefont {Cao},\ and\ \citenamefont
  {Christodoulides}}]{PhysRevLett.106.213901}%
  \BibitemOpen
  \bibfield  {author} {\bibinfo {author} {\bibfnamefont {Z.}~\bibnamefont
  {Lin}}, \bibinfo {author} {\bibfnamefont {H.}~\bibnamefont {Ramezani}},
  \bibinfo {author} {\bibfnamefont {T.}~\bibnamefont {Eichelkraut}}, \bibinfo
  {author} {\bibfnamefont {T.}~\bibnamefont {Kottos}}, \bibinfo {author}
  {\bibfnamefont {H.}~\bibnamefont {Cao}},\ and\ \bibinfo {author}
  {\bibfnamefont {D.~N.}\ \bibnamefont {Christodoulides}},\ }\bibfield  {title}
  {\bibinfo {title} {Unidirectional invisibility induced by
  $\mathcal{P}\mathcal{T}$-symmetric periodic structures},\ }\href
  {https://doi.org/10.1103/PhysRevLett.106.213901} {\bibfield  {journal}
  {\bibinfo  {journal} {Phys. Rev. Lett.}\ }\textbf {\bibinfo {volume} {106}},\
  \bibinfo {pages} {213901} (\bibinfo {year} {2011})}\BibitemShut {NoStop}%
\bibitem [{\citenamefont {{\"{O}}zdemir}\ \emph {et~al.}(2019)\citenamefont
  {{\"{O}}zdemir}, \citenamefont {Rotter}, \citenamefont {Nori},\ and\
  \citenamefont {Yang}}]{natmater.18.8}%
  \BibitemOpen
  \bibfield  {author} {\bibinfo {author} {\bibfnamefont {{\c{S}}.~K.}\
  \bibnamefont {{\"{O}}zdemir}}, \bibinfo {author} {\bibfnamefont
  {S.}~\bibnamefont {Rotter}}, \bibinfo {author} {\bibfnamefont
  {F.}~\bibnamefont {Nori}},\ and\ \bibinfo {author} {\bibfnamefont
  {L.}~\bibnamefont {Yang}},\ }\bibfield  {title} {\bibinfo {title}
  {Parity-time symmetry and exceptional points in photonics},\ }\href
  {https://doi.org/10.1038/s41563-019-0304-9} {\bibfield  {journal} {\bibinfo
  {journal} {Nat. Mater.}\ }\textbf {\bibinfo {volume} {18}},\ \bibinfo {pages}
  {783} (\bibinfo {year} {2019})}\BibitemShut {NoStop}%
\bibitem [{\citenamefont {Jalas}\ \emph {et~al.}(2013)\citenamefont {Jalas},
  \citenamefont {Petrov}, \citenamefont {Eich}, \citenamefont {Freude},
  \citenamefont {Fan}, \citenamefont {Yu}, \citenamefont {Baets}, \citenamefont
  {Popovi\'{c}}, \citenamefont {Melloni}, \citenamefont {Joannopoulos},
  \citenamefont {Vanwolleghem}, \citenamefont {Doerr},\ and\ \citenamefont
  {Renner}}]{nphoton.2013.185}%
  \BibitemOpen
  \bibfield  {author} {\bibinfo {author} {\bibfnamefont {D.}~\bibnamefont
  {Jalas}}, \bibinfo {author} {\bibfnamefont {A.}~\bibnamefont {Petrov}},
  \bibinfo {author} {\bibfnamefont {M.}~\bibnamefont {Eich}}, \bibinfo {author}
  {\bibfnamefont {W.}~\bibnamefont {Freude}}, \bibinfo {author} {\bibfnamefont
  {S.}~\bibnamefont {Fan}}, \bibinfo {author} {\bibfnamefont {Z.}~\bibnamefont
  {Yu}}, \bibinfo {author} {\bibfnamefont {R.}~\bibnamefont {Baets}}, \bibinfo
  {author} {\bibfnamefont {M.}~\bibnamefont {Popovi\'{c}}}, \bibinfo {author}
  {\bibfnamefont {A.}~\bibnamefont {Melloni}}, \bibinfo {author} {\bibfnamefont
  {J.~D.}\ \bibnamefont {Joannopoulos}}, \bibinfo {author} {\bibfnamefont
  {M.}~\bibnamefont {Vanwolleghem}}, \bibinfo {author} {\bibfnamefont {C.~R.}\
  \bibnamefont {Doerr}},\ and\ \bibinfo {author} {\bibfnamefont
  {H.}~\bibnamefont {Renner}},\ }\bibfield  {title} {\bibinfo {title} {What
  is---and what is not---an optical isolator},\ }\href
  {https://doi.org/10.1038/nphoton.2013.185} {\bibfield  {journal} {\bibinfo
  {journal} {Nat. Photonics}\ }\textbf {\bibinfo {volume} {7}},\ \bibinfo
  {pages} {579} (\bibinfo {year} {2013})}\BibitemShut {NoStop}%
\bibitem [{\citenamefont {Bahari}\ \emph {et~al.}(2017)\citenamefont {Bahari},
  \citenamefont {Ndao}, \citenamefont {Vallini}, \citenamefont {El~Amili},
  \citenamefont {Fainman},\ and\ \citenamefont {Kant\'{e}}}]{science.aao4551}%
  \BibitemOpen
  \bibfield  {author} {\bibinfo {author} {\bibfnamefont {B.}~\bibnamefont
  {Bahari}}, \bibinfo {author} {\bibfnamefont {A.}~\bibnamefont {Ndao}},
  \bibinfo {author} {\bibfnamefont {F.}~\bibnamefont {Vallini}}, \bibinfo
  {author} {\bibfnamefont {A.}~\bibnamefont {El~Amili}}, \bibinfo {author}
  {\bibfnamefont {Y.}~\bibnamefont {Fainman}},\ and\ \bibinfo {author}
  {\bibfnamefont {B.}~\bibnamefont {Kant\'{e}}},\ }\bibfield  {title} {\bibinfo
  {title} {Nonreciprocal lasing in topological cavities of arbitrary
  geometries},\ }\href {https://doi.org/10.1126/science.aao4551} {\bibfield
  {journal} {\bibinfo  {journal} {Science}\ }\textbf {\bibinfo {volume}
  {358}},\ \bibinfo {pages} {636} (\bibinfo {year} {2017})}\BibitemShut
  {NoStop}%
\bibitem [{\citenamefont {Jiang}\ \emph {et~al.}(2018)\citenamefont {Jiang},
  \citenamefont {Maayani}, \citenamefont {Carmon}, \citenamefont {Nori},\ and\
  \citenamefont {Jing}}]{PhysRevApplied.10.064037}%
  \BibitemOpen
  \bibfield  {author} {\bibinfo {author} {\bibfnamefont {Y.}~\bibnamefont
  {Jiang}}, \bibinfo {author} {\bibfnamefont {S.}~\bibnamefont {Maayani}},
  \bibinfo {author} {\bibfnamefont {T.}~\bibnamefont {Carmon}}, \bibinfo
  {author} {\bibfnamefont {F.}~\bibnamefont {Nori}},\ and\ \bibinfo {author}
  {\bibfnamefont {H.}~\bibnamefont {Jing}},\ }\bibfield  {title} {\bibinfo
  {title} {Nonreciprocal phonon laser},\ }\href
  {https://doi.org/10.1103/PhysRevApplied.10.064037} {\bibfield  {journal}
  {\bibinfo  {journal} {Phys. Rev. Appl.}\ }\textbf {\bibinfo {volume} {10}},\
  \bibinfo {pages} {064037} (\bibinfo {year} {2018})}\BibitemShut {NoStop}%
\bibitem [{\citenamefont {Peng}\ \emph {et~al.}(2014)\citenamefont {Peng},
  \citenamefont {{\"{O}}zdemir}, \citenamefont {Lei}, \citenamefont {Monifi},
  \citenamefont {Gianfreda}, \citenamefont {Long}, \citenamefont {Fan},
  \citenamefont {Nori}, \citenamefont {Bender},\ and\ \citenamefont
  {Yang}}]{nphysics.10.5}%
  \BibitemOpen
  \bibfield  {author} {\bibinfo {author} {\bibfnamefont {B.}~\bibnamefont
  {Peng}}, \bibinfo {author} {\bibfnamefont {{\c{S}}.~K.}\ \bibnamefont
  {{\"{O}}zdemir}}, \bibinfo {author} {\bibfnamefont {F.}~\bibnamefont {Lei}},
  \bibinfo {author} {\bibfnamefont {F.}~\bibnamefont {Monifi}}, \bibinfo
  {author} {\bibfnamefont {M.}~\bibnamefont {Gianfreda}}, \bibinfo {author}
  {\bibfnamefont {G.~L.}\ \bibnamefont {Long}}, \bibinfo {author}
  {\bibfnamefont {S.}~\bibnamefont {Fan}}, \bibinfo {author} {\bibfnamefont
  {F.}~\bibnamefont {Nori}}, \bibinfo {author} {\bibfnamefont {C.~M.}\
  \bibnamefont {Bender}},\ and\ \bibinfo {author} {\bibfnamefont
  {L.}~\bibnamefont {Yang}},\ }\bibfield  {title} {\bibinfo {title}
  {Parity-time-symmetric whispering-gallery microcavities},\ }\href
  {https://doi.org/10.1038/nphys2927} {\bibfield  {journal} {\bibinfo
  {journal} {Nat. Phys.}\ }\textbf {\bibinfo {volume} {10}},\ \bibinfo {pages}
  {394} (\bibinfo {year} {2014})}\BibitemShut {NoStop}%
\bibitem [{\citenamefont {Cao}\ \emph {et~al.}(2017)\citenamefont {Cao},
  \citenamefont {Wang}, \citenamefont {Dong}, \citenamefont {Jing},
  \citenamefont {Liu}, \citenamefont {Chen}, \citenamefont {Ge}, \citenamefont
  {Gong},\ and\ \citenamefont {Xiao}}]{PhysRevLett.118.033901}%
  \BibitemOpen
  \bibfield  {author} {\bibinfo {author} {\bibfnamefont {Q.-T.}\ \bibnamefont
  {Cao}}, \bibinfo {author} {\bibfnamefont {H.}~\bibnamefont {Wang}}, \bibinfo
  {author} {\bibfnamefont {C.-H.}\ \bibnamefont {Dong}}, \bibinfo {author}
  {\bibfnamefont {H.}~\bibnamefont {Jing}}, \bibinfo {author} {\bibfnamefont
  {R.-S.}\ \bibnamefont {Liu}}, \bibinfo {author} {\bibfnamefont
  {X.}~\bibnamefont {Chen}}, \bibinfo {author} {\bibfnamefont {L.}~\bibnamefont
  {Ge}}, \bibinfo {author} {\bibfnamefont {Q.}~\bibnamefont {Gong}},\ and\
  \bibinfo {author} {\bibfnamefont {Y.-F.}\ \bibnamefont {Xiao}},\ }\bibfield
  {title} {\bibinfo {title} {Experimental demonstration of spontaneous
  chirality in a nonlinear microresonator},\ }\href
  {https://doi.org/10.1103/PhysRevLett.118.033901} {\bibfield  {journal}
  {\bibinfo  {journal} {Phys. Rev. Lett.}\ }\textbf {\bibinfo {volume} {118}},\
  \bibinfo {pages} {033901} (\bibinfo {year} {2017})}\BibitemShut {NoStop}%
\bibitem [{\citenamefont {Zhang}\ \emph {et~al.}(2018)\citenamefont {Zhang},
  \citenamefont {Hu}, \citenamefont {Lin}, \citenamefont {Niu}, \citenamefont
  {Xia}, \citenamefont {Gong},\ and\ \citenamefont {Gong}}]{nphoton.2018.1038}%
  \BibitemOpen
  \bibfield  {author} {\bibinfo {author} {\bibfnamefont {S.}~\bibnamefont
  {Zhang}}, \bibinfo {author} {\bibfnamefont {Y.}~\bibnamefont {Hu}}, \bibinfo
  {author} {\bibfnamefont {G.}~\bibnamefont {Lin}}, \bibinfo {author}
  {\bibfnamefont {Y.}~\bibnamefont {Niu}}, \bibinfo {author} {\bibfnamefont
  {K.}~\bibnamefont {Xia}}, \bibinfo {author} {\bibfnamefont {J.}~\bibnamefont
  {Gong}},\ and\ \bibinfo {author} {\bibfnamefont {S.}~\bibnamefont {Gong}},\
  }\bibfield  {title} {\bibinfo {title} {Thermal-motion-induced non-reciprocal
  quantum optical system},\ }\href {https://doi.org/10.1038/s41566-018-0269-2}
  {\bibfield  {journal} {\bibinfo  {journal} {Nat. Photonics}\ }\textbf
  {\bibinfo {volume} {12}},\ \bibinfo {pages} {744} (\bibinfo {year}
  {2018})}\BibitemShut {NoStop}%
\bibitem [{\citenamefont {Xia}\ \emph {et~al.}(2018)\citenamefont {Xia},
  \citenamefont {Nori},\ and\ \citenamefont {Xiao}}]{PhysRevLett.121.203602}%
  \BibitemOpen
  \bibfield  {author} {\bibinfo {author} {\bibfnamefont {K.}~\bibnamefont
  {Xia}}, \bibinfo {author} {\bibfnamefont {F.}~\bibnamefont {Nori}},\ and\
  \bibinfo {author} {\bibfnamefont {M.}~\bibnamefont {Xiao}},\ }\bibfield
  {title} {\bibinfo {title} {Cavity-free optical isolators and circulators
  using a chiral cross-{Kerr} nonlinearity},\ }\href
  {https://doi.org/10.1103/PhysRevLett.121.203602} {\bibfield  {journal}
  {\bibinfo  {journal} {Phys. Rev. Lett.}\ }\textbf {\bibinfo {volume} {121}},\
  \bibinfo {pages} {203602} (\bibinfo {year} {2018})}\BibitemShut {NoStop}%
\bibitem [{\citenamefont {Yang}\ \emph {et~al.}(2019)\citenamefont {Yang},
  \citenamefont {Xia}, \citenamefont {He}, \citenamefont {Li}, \citenamefont
  {Han}, \citenamefont {Zhang}, \citenamefont {Li}, \citenamefont {Zhang},
  \citenamefont {Xu}, \citenamefont {Yang},\ and\ \citenamefont
  {Zhang}}]{PhysRevLett.123.233604}%
  \BibitemOpen
  \bibfield  {author} {\bibinfo {author} {\bibfnamefont {P.}~\bibnamefont
  {Yang}}, \bibinfo {author} {\bibfnamefont {X.}~\bibnamefont {Xia}}, \bibinfo
  {author} {\bibfnamefont {H.}~\bibnamefont {He}}, \bibinfo {author}
  {\bibfnamefont {S.}~\bibnamefont {Li}}, \bibinfo {author} {\bibfnamefont
  {X.}~\bibnamefont {Han}}, \bibinfo {author} {\bibfnamefont {P.}~\bibnamefont
  {Zhang}}, \bibinfo {author} {\bibfnamefont {G.}~\bibnamefont {Li}}, \bibinfo
  {author} {\bibfnamefont {P.}~\bibnamefont {Zhang}}, \bibinfo {author}
  {\bibfnamefont {J.}~\bibnamefont {Xu}}, \bibinfo {author} {\bibfnamefont
  {Y.}~\bibnamefont {Yang}},\ and\ \bibinfo {author} {\bibfnamefont
  {T.}~\bibnamefont {Zhang}},\ }\bibfield  {title} {\bibinfo {title}
  {Realization of nonlinear optical nonreciprocity on a few-photon level based
  on atoms strongly coupled to an asymmetric cavity},\ }\href
  {https://doi.org/10.1103/PhysRevLett.123.233604} {\bibfield  {journal}
  {\bibinfo  {journal} {Phys. Rev. Lett.}\ }\textbf {\bibinfo {volume} {123}},\
  \bibinfo {pages} {233604} (\bibinfo {year} {2019})}\BibitemShut {NoStop}%
\bibitem [{\citenamefont {Li}\ \emph {et~al.}(2020)\citenamefont {Li},
  \citenamefont {Ding}, \citenamefont {Yu}, \citenamefont {Dong}, \citenamefont
  {Zeng}, \citenamefont {Zhang}, \citenamefont {Ye}, \citenamefont {Wu},
  \citenamefont {Zhu}, \citenamefont {Gao}, \citenamefont {Guo},\ and\
  \citenamefont {Shi}}]{PhysRevResearch.2.033517}%
  \BibitemOpen
  \bibfield  {author} {\bibinfo {author} {\bibfnamefont {E.-Z.}\ \bibnamefont
  {Li}}, \bibinfo {author} {\bibfnamefont {D.-S.}\ \bibnamefont {Ding}},
  \bibinfo {author} {\bibfnamefont {Y.-C.}\ \bibnamefont {Yu}}, \bibinfo
  {author} {\bibfnamefont {M.-X.}\ \bibnamefont {Dong}}, \bibinfo {author}
  {\bibfnamefont {L.}~\bibnamefont {Zeng}}, \bibinfo {author} {\bibfnamefont
  {W.-H.}\ \bibnamefont {Zhang}}, \bibinfo {author} {\bibfnamefont {Y.-H.}\
  \bibnamefont {Ye}}, \bibinfo {author} {\bibfnamefont {H.-Z.}\ \bibnamefont
  {Wu}}, \bibinfo {author} {\bibfnamefont {Z.-H.}\ \bibnamefont {Zhu}},
  \bibinfo {author} {\bibfnamefont {W.}~\bibnamefont {Gao}}, \bibinfo {author}
  {\bibfnamefont {G.-C.}\ \bibnamefont {Guo}},\ and\ \bibinfo {author}
  {\bibfnamefont {B.-S.}\ \bibnamefont {Shi}},\ }\bibfield  {title} {\bibinfo
  {title} {Experimental demonstration of cavity-free optical isolators and
  optical circulators},\ }\href
  {https://doi.org/10.1103/PhysRevResearch.2.033517} {\bibfield  {journal}
  {\bibinfo  {journal} {Phys. Rev. Res.}\ }\textbf {\bibinfo {volume} {2}},\
  \bibinfo {pages} {033517} (\bibinfo {year} {2020})}\BibitemShut {NoStop}%
\bibitem [{\citenamefont {Tang}\ \emph
  {et~al.}(2021{\natexlab{a}})\citenamefont {Tang}, \citenamefont {Tang},
  \citenamefont {Wu}, \citenamefont {Zhang}, \citenamefont {Xiao},\ and\
  \citenamefont {Xia}}]{PRJ.413286}%
  \BibitemOpen
  \bibfield  {author} {\bibinfo {author} {\bibfnamefont {L.}~\bibnamefont
  {Tang}}, \bibinfo {author} {\bibfnamefont {J.}~\bibnamefont {Tang}}, \bibinfo
  {author} {\bibfnamefont {H.}~\bibnamefont {Wu}}, \bibinfo {author}
  {\bibfnamefont {J.}~\bibnamefont {Zhang}}, \bibinfo {author} {\bibfnamefont
  {M.}~\bibnamefont {Xiao}},\ and\ \bibinfo {author} {\bibfnamefont
  {K.}~\bibnamefont {Xia}},\ }\bibfield  {title} {\bibinfo {title}
  {Broad-intensity-range optical nonreciprocity based on feedback-induced
  {Kerr} nonlinearity},\ }\href {https://doi.org/10.1364/PRJ.413286} {\bibfield
   {journal} {\bibinfo  {journal} {Photonics Res.}\ }\textbf {\bibinfo {volume}
  {9}},\ \bibinfo {pages} {1218} (\bibinfo {year}
  {2021}{\natexlab{a}})}\BibitemShut {NoStop}%
\bibitem [{\citenamefont {Tang}\ \emph
  {et~al.}(2021{\natexlab{b}})\citenamefont {Tang}, \citenamefont {Tang},
  \citenamefont {Chen}, \citenamefont {Nori}, \citenamefont {Xiao},\ and\
  \citenamefont {Xia}}]{tanglei.3970886}%
  \BibitemOpen
  \bibfield  {author} {\bibinfo {author} {\bibfnamefont {L.}~\bibnamefont
  {Tang}}, \bibinfo {author} {\bibfnamefont {J.}~\bibnamefont {Tang}}, \bibinfo
  {author} {\bibfnamefont {M.}~\bibnamefont {Chen}}, \bibinfo {author}
  {\bibfnamefont {F.}~\bibnamefont {Nori}}, \bibinfo {author} {\bibfnamefont
  {M.}~\bibnamefont {Xiao}},\ and\ \bibinfo {author} {\bibfnamefont
  {K.}~\bibnamefont {Xia}},\ }\bibfield  {title} {\bibinfo {title} {Quantum
  squeezing induced optical nonreciprocity},\ }\href
  {https://arxiv.org/abs/2110.05016} {\bibfield  {journal} {\bibinfo  {journal}
  {arXiv:2110.05016 [quant-ph]}\ } (\bibinfo {year}
  {2021}{\natexlab{b}})}\BibitemShut {NoStop}%
\bibitem [{\citenamefont {Hafezi}\ and\ \citenamefont
  {Rabl}(2012)}]{OE.20.007672}%
  \BibitemOpen
  \bibfield  {author} {\bibinfo {author} {\bibfnamefont {M.}~\bibnamefont
  {Hafezi}}\ and\ \bibinfo {author} {\bibfnamefont {P.}~\bibnamefont {Rabl}},\
  }\bibfield  {title} {\bibinfo {title} {Optomechanically induced
  non-reciprocity in microring resonators},\ }\href
  {https://doi.org/10.1364/OE.20.007672} {\bibfield  {journal} {\bibinfo
  {journal} {Opt. Express}\ }\textbf {\bibinfo {volume} {20}},\ \bibinfo
  {pages} {7672} (\bibinfo {year} {2012})}\BibitemShut {NoStop}%
\bibitem [{\citenamefont {Shen}\ \emph {et~al.}(2016)\citenamefont {Shen},
  \citenamefont {Zhang}, \citenamefont {Chen}, \citenamefont {Zou},
  \citenamefont {Xiao}, \citenamefont {Zou}, \citenamefont {Sun}, \citenamefont
  {Guo},\ and\ \citenamefont {Dong}}]{nphoton.2016.161}%
  \BibitemOpen
  \bibfield  {author} {\bibinfo {author} {\bibfnamefont {Z.}~\bibnamefont
  {Shen}}, \bibinfo {author} {\bibfnamefont {Y.-L.}\ \bibnamefont {Zhang}},
  \bibinfo {author} {\bibfnamefont {Y.}~\bibnamefont {Chen}}, \bibinfo {author}
  {\bibfnamefont {C.-L.}\ \bibnamefont {Zou}}, \bibinfo {author} {\bibfnamefont
  {Y.-F.}\ \bibnamefont {Xiao}}, \bibinfo {author} {\bibfnamefont {X.-B.}\
  \bibnamefont {Zou}}, \bibinfo {author} {\bibfnamefont {F.-W.}\ \bibnamefont
  {Sun}}, \bibinfo {author} {\bibfnamefont {G.-C.}\ \bibnamefont {Guo}},\ and\
  \bibinfo {author} {\bibfnamefont {C.-H.}\ \bibnamefont {Dong}},\ }\bibfield
  {title} {\bibinfo {title} {Experimental realization of optomechanically
  induced non-reciprocity},\ }\href {https://doi.org/10.1038/nphoton.2016.161}
  {\bibfield  {journal} {\bibinfo  {journal} {Nat. Photonics}\ }\textbf
  {\bibinfo {volume} {10}},\ \bibinfo {pages} {657} (\bibinfo {year}
  {2016})}\BibitemShut {NoStop}%
\bibitem [{\citenamefont {Ruesink}\ \emph {et~al.}(2016)\citenamefont
  {Ruesink}, \citenamefont {Miri}, \citenamefont {Al\`{u}},\ and\ \citenamefont
  {Verhagen}}]{ncomms13662}%
  \BibitemOpen
  \bibfield  {author} {\bibinfo {author} {\bibfnamefont {F.}~\bibnamefont
  {Ruesink}}, \bibinfo {author} {\bibfnamefont {M.-A.}\ \bibnamefont {Miri}},
  \bibinfo {author} {\bibfnamefont {A.}~\bibnamefont {Al\`{u}}},\ and\ \bibinfo
  {author} {\bibfnamefont {E.}~\bibnamefont {Verhagen}},\ }\bibfield  {title}
  {\bibinfo {title} {Nonreciprocity and magnetic-free isolation based on
  optomechanical interactions},\ }\href {https://doi.org/10.1038/ncomms13662}
  {\bibfield  {journal} {\bibinfo  {journal} {Nat. Commun.}\ }\textbf {\bibinfo
  {volume} {7}},\ \bibinfo {pages} {13662} (\bibinfo {year}
  {2016})}\BibitemShut {NoStop}%
\bibitem [{\citenamefont {Maayani}\ \emph {et~al.}(2018)\citenamefont
  {Maayani}, \citenamefont {Dahan}, \citenamefont {Kligerman}, \citenamefont
  {Moses}, \citenamefont {Hassan}, \citenamefont {Jing}, \citenamefont {Nori},
  \citenamefont {Christodoulides},\ and\ \citenamefont
  {Carmon}}]{natures41586.018}%
  \BibitemOpen
  \bibfield  {author} {\bibinfo {author} {\bibfnamefont {S.}~\bibnamefont
  {Maayani}}, \bibinfo {author} {\bibfnamefont {R.}~\bibnamefont {Dahan}},
  \bibinfo {author} {\bibfnamefont {Y.}~\bibnamefont {Kligerman}}, \bibinfo
  {author} {\bibfnamefont {E.}~\bibnamefont {Moses}}, \bibinfo {author}
  {\bibfnamefont {A.~U.}\ \bibnamefont {Hassan}}, \bibinfo {author}
  {\bibfnamefont {H.}~\bibnamefont {Jing}}, \bibinfo {author} {\bibfnamefont
  {F.}~\bibnamefont {Nori}}, \bibinfo {author} {\bibfnamefont {D.~N.}\
  \bibnamefont {Christodoulides}},\ and\ \bibinfo {author} {\bibfnamefont
  {T.}~\bibnamefont {Carmon}},\ }\bibfield  {title} {\bibinfo {title} {Flying
  couplers above spinning resonators generate irreversible refraction},\ }\href
  {https://doi.org/10.1038/s41586-018-0245-5} {\bibfield  {journal} {\bibinfo
  {journal} {Nature (London)}\ }\textbf {\bibinfo {volume} {558}},\ \bibinfo
  {pages} {569} (\bibinfo {year} {2018})}\BibitemShut {NoStop}%
\bibitem [{\citenamefont {Huang}\ \emph {et~al.}(2018)\citenamefont {Huang},
  \citenamefont {Miranowicz}, \citenamefont {Liao}, \citenamefont {Nori},\ and\
  \citenamefont {Jing}}]{PhysRevLett.121.153601}%
  \BibitemOpen
  \bibfield  {author} {\bibinfo {author} {\bibfnamefont {R.}~\bibnamefont
  {Huang}}, \bibinfo {author} {\bibfnamefont {A.}~\bibnamefont {Miranowicz}},
  \bibinfo {author} {\bibfnamefont {J.-Q.}\ \bibnamefont {Liao}}, \bibinfo
  {author} {\bibfnamefont {F.}~\bibnamefont {Nori}},\ and\ \bibinfo {author}
  {\bibfnamefont {H.}~\bibnamefont {Jing}},\ }\bibfield  {title} {\bibinfo
  {title} {Nonreciprocal photon blockade},\ }\href
  {https://doi.org/10.1103/PhysRevLett.121.153601} {\bibfield  {journal}
  {\bibinfo  {journal} {Phys. Rev. Lett.}\ }\textbf {\bibinfo {volume} {121}},\
  \bibinfo {pages} {153601} (\bibinfo {year} {2018})}\BibitemShut {NoStop}%
\bibitem [{\citenamefont {Wang}\ \emph {et~al.}(2013)\citenamefont {Wang},
  \citenamefont {Zhou}, \citenamefont {Guo}, \citenamefont {Zhang},
  \citenamefont {Evers},\ and\ \citenamefont {Zhu}}]{PhysRevLett.110.093901}%
  \BibitemOpen
  \bibfield  {author} {\bibinfo {author} {\bibfnamefont {D.-W.}\ \bibnamefont
  {Wang}}, \bibinfo {author} {\bibfnamefont {H.-T.}\ \bibnamefont {Zhou}},
  \bibinfo {author} {\bibfnamefont {M.-J.}\ \bibnamefont {Guo}}, \bibinfo
  {author} {\bibfnamefont {J.-X.}\ \bibnamefont {Zhang}}, \bibinfo {author}
  {\bibfnamefont {J.}~\bibnamefont {Evers}},\ and\ \bibinfo {author}
  {\bibfnamefont {S.-Y.}\ \bibnamefont {Zhu}},\ }\bibfield  {title} {\bibinfo
  {title} {Optical diode made from a moving photonic crystal},\ }\href
  {https://doi.org/10.1103/PhysRevLett.110.093901} {\bibfield  {journal}
  {\bibinfo  {journal} {Phys. Rev. Lett.}\ }\textbf {\bibinfo {volume} {110}},\
  \bibinfo {pages} {093901} (\bibinfo {year} {2013})}\BibitemShut {NoStop}%
\bibitem [{\citenamefont {Horsley}\ \emph {et~al.}(2013)\citenamefont
  {Horsley}, \citenamefont {Wu}, \citenamefont {Artoni},\ and\ \citenamefont
  {La~Rocca}}]{PhysRevLett.110.223602}%
  \BibitemOpen
  \bibfield  {author} {\bibinfo {author} {\bibfnamefont {S.~A.~R.}\
  \bibnamefont {Horsley}}, \bibinfo {author} {\bibfnamefont {J.-H.}\
  \bibnamefont {Wu}}, \bibinfo {author} {\bibfnamefont {M.}~\bibnamefont
  {Artoni}},\ and\ \bibinfo {author} {\bibfnamefont {G.~C.}\ \bibnamefont
  {La~Rocca}},\ }\bibfield  {title} {\bibinfo {title} {Optical nonreciprocity
  of cold atom {Bragg} mirrors in motion},\ }\href
  {https://doi.org/10.1103/PhysRevLett.110.223602} {\bibfield  {journal}
  {\bibinfo  {journal} {Phys. Rev. Lett.}\ }\textbf {\bibinfo {volume} {110}},\
  \bibinfo {pages} {223602} (\bibinfo {year} {2013})}\BibitemShut {NoStop}%
\bibitem [{\citenamefont {Sounas}\ and\ \citenamefont
  {Al\`{u}}(2017)}]{nphotons41566.017}%
  \BibitemOpen
  \bibfield  {author} {\bibinfo {author} {\bibfnamefont {D.~L.}\ \bibnamefont
  {Sounas}}\ and\ \bibinfo {author} {\bibfnamefont {A.}~\bibnamefont
  {Al\`{u}}},\ }\bibfield  {title} {\bibinfo {title} {Non-reciprocal photonics
  based on time modulation},\ }\href
  {https://doi.org/10.1038/s41566-017-0051-x} {\bibfield  {journal} {\bibinfo
  {journal} {Nat. Photonics}\ }\textbf {\bibinfo {volume} {11}},\ \bibinfo
  {pages} {774} (\bibinfo {year} {2017})}\BibitemShut {NoStop}%
\bibitem [{\citenamefont {Sayrin}\ \emph {et~al.}(2015)\citenamefont {Sayrin},
  \citenamefont {Junge}, \citenamefont {Mitsch}, \citenamefont {Albrecht},
  \citenamefont {O'Shea}, \citenamefont {Schneeweiss}, \citenamefont {Volz},\
  and\ \citenamefont {Rauschenbeutel}}]{PhysRevX.5.041036}%
  \BibitemOpen
  \bibfield  {author} {\bibinfo {author} {\bibfnamefont {C.}~\bibnamefont
  {Sayrin}}, \bibinfo {author} {\bibfnamefont {C.}~\bibnamefont {Junge}},
  \bibinfo {author} {\bibfnamefont {R.}~\bibnamefont {Mitsch}}, \bibinfo
  {author} {\bibfnamefont {B.}~\bibnamefont {Albrecht}}, \bibinfo {author}
  {\bibfnamefont {D.}~\bibnamefont {O'Shea}}, \bibinfo {author} {\bibfnamefont
  {P.}~\bibnamefont {Schneeweiss}}, \bibinfo {author} {\bibfnamefont
  {J.}~\bibnamefont {Volz}},\ and\ \bibinfo {author} {\bibfnamefont
  {A.}~\bibnamefont {Rauschenbeutel}},\ }\bibfield  {title} {\bibinfo {title}
  {Nanophotonic optical isolator controlled by the internal state of cold
  atoms},\ }\href {https://doi.org/10.1103/PhysRevX.5.041036} {\bibfield
  {journal} {\bibinfo  {journal} {Phys. Rev. X}\ }\textbf {\bibinfo {volume}
  {5}},\ \bibinfo {pages} {041036} (\bibinfo {year} {2015})}\BibitemShut
  {NoStop}%
\bibitem [{\citenamefont {Xia}\ \emph {et~al.}(2014)\citenamefont {Xia},
  \citenamefont {Lu}, \citenamefont {Lin}, \citenamefont {Cheng}, \citenamefont
  {Niu}, \citenamefont {Gong},\ and\ \citenamefont
  {Twamley}}]{PhysRevA.90.043802}%
  \BibitemOpen
  \bibfield  {author} {\bibinfo {author} {\bibfnamefont {K.}~\bibnamefont
  {Xia}}, \bibinfo {author} {\bibfnamefont {G.}~\bibnamefont {Lu}}, \bibinfo
  {author} {\bibfnamefont {G.}~\bibnamefont {Lin}}, \bibinfo {author}
  {\bibfnamefont {Y.}~\bibnamefont {Cheng}}, \bibinfo {author} {\bibfnamefont
  {Y.}~\bibnamefont {Niu}}, \bibinfo {author} {\bibfnamefont {S.}~\bibnamefont
  {Gong}},\ and\ \bibinfo {author} {\bibfnamefont {J.}~\bibnamefont
  {Twamley}},\ }\bibfield  {title} {\bibinfo {title} {Reversible nonmagnetic
  single-photon isolation using unbalanced quantum coupling},\ }\href
  {https://doi.org/10.1103/PhysRevA.90.043802} {\bibfield  {journal} {\bibinfo
  {journal} {Phys. Rev. A}\ }\textbf {\bibinfo {volume} {90}},\ \bibinfo
  {pages} {043802} (\bibinfo {year} {2014})}\BibitemShut {NoStop}%
\bibitem [{\citenamefont {Scheucher}\ \emph {et~al.}(2016)\citenamefont
  {Scheucher}, \citenamefont {Hilico}, \citenamefont {Will}, \citenamefont
  {Volz},\ and\ \citenamefont {Rauschenbeutel}}]{science.aaj2118}%
  \BibitemOpen
  \bibfield  {author} {\bibinfo {author} {\bibfnamefont {M.}~\bibnamefont
  {Scheucher}}, \bibinfo {author} {\bibfnamefont {A.}~\bibnamefont {Hilico}},
  \bibinfo {author} {\bibfnamefont {E.}~\bibnamefont {Will}}, \bibinfo {author}
  {\bibfnamefont {J.}~\bibnamefont {Volz}},\ and\ \bibinfo {author}
  {\bibfnamefont {A.}~\bibnamefont {Rauschenbeutel}},\ }\bibfield  {title}
  {\bibinfo {title} {Quantum optical circulator controlled by a single chirally
  coupled atom},\ }\href {https://doi.org/10.1126/science.aaj2118} {\bibfield
  {journal} {\bibinfo  {journal} {Science}\ }\textbf {\bibinfo {volume}
  {354}},\ \bibinfo {pages} {1577} (\bibinfo {year} {2016})}\BibitemShut
  {NoStop}%
\bibitem [{\citenamefont {Lodahl}\ \emph {et~al.}(2017)\citenamefont {Lodahl},
  \citenamefont {Mahmoodian}, \citenamefont {Stobbe}, \citenamefont
  {Rauschenbeutel}, \citenamefont {Schneeweiss}, \citenamefont {Volz},
  \citenamefont {Pichler},\ and\ \citenamefont {Zoller}}]{nature21037}%
  \BibitemOpen
  \bibfield  {author} {\bibinfo {author} {\bibfnamefont {P.}~\bibnamefont
  {Lodahl}}, \bibinfo {author} {\bibfnamefont {S.}~\bibnamefont {Mahmoodian}},
  \bibinfo {author} {\bibfnamefont {S.}~\bibnamefont {Stobbe}}, \bibinfo
  {author} {\bibfnamefont {A.}~\bibnamefont {Rauschenbeutel}}, \bibinfo
  {author} {\bibfnamefont {P.}~\bibnamefont {Schneeweiss}}, \bibinfo {author}
  {\bibfnamefont {J.}~\bibnamefont {Volz}}, \bibinfo {author} {\bibfnamefont
  {H.}~\bibnamefont {Pichler}},\ and\ \bibinfo {author} {\bibfnamefont
  {P.}~\bibnamefont {Zoller}},\ }\bibfield  {title} {\bibinfo {title} {Chiral
  quantum optics},\ }\href {https://doi.org/10.1038/nature21037} {\bibfield
  {journal} {\bibinfo  {journal} {Nature (London)}\ }\textbf {\bibinfo {volume}
  {541}},\ \bibinfo {pages} {473} (\bibinfo {year} {2017})}\BibitemShut
  {NoStop}%
\bibitem [{\citenamefont {Mehranad}\ \emph {et~al.}(2020)\citenamefont
  {Mehranad}, \citenamefont {Foster}, \citenamefont {Dost}, \citenamefont
  {Clarke}, \citenamefont {Patil}, \citenamefont {Fox}, \citenamefont
  {Skolnick},\ and\ \citenamefont {Wilson}}]{OPTICA.393035}%
  \BibitemOpen
  \bibfield  {author} {\bibinfo {author} {\bibfnamefont {M.~J.}\ \bibnamefont
  {Mehranad}}, \bibinfo {author} {\bibfnamefont {A.~P.}\ \bibnamefont
  {Foster}}, \bibinfo {author} {\bibfnamefont {R.}~\bibnamefont {Dost}},
  \bibinfo {author} {\bibfnamefont {E.}~\bibnamefont {Clarke}}, \bibinfo
  {author} {\bibfnamefont {P.~K.}\ \bibnamefont {Patil}}, \bibinfo {author}
  {\bibfnamefont {A.~M.}\ \bibnamefont {Fox}}, \bibinfo {author} {\bibfnamefont
  {M.~S.}\ \bibnamefont {Skolnick}},\ and\ \bibinfo {author} {\bibfnamefont
  {L.~R.}\ \bibnamefont {Wilson}},\ }\bibfield  {title} {\bibinfo {title}
  {Chiral topological photonics with an embedded quantum emitter},\ }\href
  {https://doi.org/10.1364/OPTICA.393035} {\bibfield  {journal} {\bibinfo
  {journal} {Optica}\ }\textbf {\bibinfo {volume} {7}},\ \bibinfo {pages}
  {1690} (\bibinfo {year} {2020})}\BibitemShut {NoStop}%
\bibitem [{\citenamefont {Dong}\ \emph {et~al.}(2021)\citenamefont {Dong},
  \citenamefont {Xia}, \citenamefont {Zhang}, \citenamefont {Yu}, \citenamefont
  {Ye}, \citenamefont {Li}, \citenamefont {Zeng}, \citenamefont {Ding},
  \citenamefont {Shi}, \citenamefont {Guo},\ and\ \citenamefont
  {Nori}}]{sciadv.abe8924}%
  \BibitemOpen
  \bibfield  {author} {\bibinfo {author} {\bibfnamefont {M.-X.}\ \bibnamefont
  {Dong}}, \bibinfo {author} {\bibfnamefont {K.}~\bibnamefont {Xia}}, \bibinfo
  {author} {\bibfnamefont {W.-H.}\ \bibnamefont {Zhang}}, \bibinfo {author}
  {\bibfnamefont {Y.-C.}\ \bibnamefont {Yu}}, \bibinfo {author} {\bibfnamefont
  {Y.-H.}\ \bibnamefont {Ye}}, \bibinfo {author} {\bibfnamefont {E.-Z.}\
  \bibnamefont {Li}}, \bibinfo {author} {\bibfnamefont {L.}~\bibnamefont
  {Zeng}}, \bibinfo {author} {\bibfnamefont {D.-S.}\ \bibnamefont {Ding}},
  \bibinfo {author} {\bibfnamefont {B.-S.}\ \bibnamefont {Shi}}, \bibinfo
  {author} {\bibfnamefont {G.-C.}\ \bibnamefont {Guo}},\ and\ \bibinfo {author}
  {\bibfnamefont {F.}~\bibnamefont {Nori}},\ }\bibfield  {title} {\bibinfo
  {title} {All-optical reversible single-photon isolation at room
  temperature},\ }\href {https://doi.org/10.1126/sciadv.abe8924} {\bibfield
  {journal} {\bibinfo  {journal} {Sci. Adv.}\ }\textbf {\bibinfo {volume}
  {7}},\ \bibinfo {pages} {eabe8924} (\bibinfo {year} {2021})}\BibitemShut
  {NoStop}%
\bibitem [{\citenamefont {Hu}\ \emph {et~al.}(2021)\citenamefont {Hu},
  \citenamefont {Wang}, \citenamefont {Zhang}, \citenamefont {Chen},
  \citenamefont {Zhang}, \citenamefont {Li}, \citenamefont {Zou}, \citenamefont
  {Zhang}, \citenamefont {Tang}, \citenamefont {Dong}, \citenamefont {Guo},\
  and\ \citenamefont {Zou}}]{ncomun.10.1038.2389}%
  \BibitemOpen
  \bibfield  {author} {\bibinfo {author} {\bibfnamefont {X.-X.}\ \bibnamefont
  {Hu}}, \bibinfo {author} {\bibfnamefont {Z.-B.}\ \bibnamefont {Wang}},
  \bibinfo {author} {\bibfnamefont {P.}~\bibnamefont {Zhang}}, \bibinfo
  {author} {\bibfnamefont {G.-J.}\ \bibnamefont {Chen}}, \bibinfo {author}
  {\bibfnamefont {Y.-L.}\ \bibnamefont {Zhang}}, \bibinfo {author}
  {\bibfnamefont {G.}~\bibnamefont {Li}}, \bibinfo {author} {\bibfnamefont
  {X.-B.}\ \bibnamefont {Zou}}, \bibinfo {author} {\bibfnamefont
  {T.}~\bibnamefont {Zhang}}, \bibinfo {author} {\bibfnamefont {H.~X.}\
  \bibnamefont {Tang}}, \bibinfo {author} {\bibfnamefont {C.-H.}\ \bibnamefont
  {Dong}}, \bibinfo {author} {\bibfnamefont {G.-C.}\ \bibnamefont {Guo}},\ and\
  \bibinfo {author} {\bibfnamefont {C.-L.}\ \bibnamefont {Zou}},\ }\bibfield
  {title} {\bibinfo {title} {Noiseless photonic non-reciprocity via
  optically-induced magnetization},\ }\href
  {https://doi.org/10.1038/s41467-021-22597-z} {\bibfield  {journal} {\bibinfo
  {journal} {Nat. Commun.}\ }\textbf {\bibinfo {volume} {12}},\ \bibinfo
  {pages} {2389} (\bibinfo {year} {2021})}\BibitemShut {NoStop}%
\bibitem [{\citenamefont {Pucher}\ \emph {et~al.}(2021)\citenamefont {Pucher},
  \citenamefont {Liedl}, \citenamefont {Jin}, \citenamefont {Rauschenbeutel},\
  and\ \citenamefont {Schneeweiss}}]{pucher2021atomic}%
  \BibitemOpen
  \bibfield  {author} {\bibinfo {author} {\bibfnamefont {S.}~\bibnamefont
  {Pucher}}, \bibinfo {author} {\bibfnamefont {C.}~\bibnamefont {Liedl}},
  \bibinfo {author} {\bibfnamefont {S.}~\bibnamefont {Jin}}, \bibinfo {author}
  {\bibfnamefont {A.}~\bibnamefont {Rauschenbeutel}},\ and\ \bibinfo {author}
  {\bibfnamefont {P.}~\bibnamefont {Schneeweiss}},\ }\bibfield  {title}
  {\bibinfo {title} {Atomic spin-controlled non-reciprocal raman amplification
  of fibre-guided light},\ }\href {https://arxiv.org/abs/2107.07272} {\bibfield
   {journal} {\bibinfo  {journal} {arXiv preprint arXiv:2107.07272}\ }
  (\bibinfo {year} {2021})}\BibitemShut {NoStop}%
\bibitem [{\citenamefont {Tang}\ \emph {et~al.}(2019)\citenamefont {Tang},
  \citenamefont {Tang}, \citenamefont {Zhang}, \citenamefont {Lu},
  \citenamefont {Zhang}, \citenamefont {Zhang}, \citenamefont {Xia},\ and\
  \citenamefont {Xiao}}]{PhysRevA.99.043833}%
  \BibitemOpen
  \bibfield  {author} {\bibinfo {author} {\bibfnamefont {L.}~\bibnamefont
  {Tang}}, \bibinfo {author} {\bibfnamefont {J.}~\bibnamefont {Tang}}, \bibinfo
  {author} {\bibfnamefont {W.}~\bibnamefont {Zhang}}, \bibinfo {author}
  {\bibfnamefont {G.}~\bibnamefont {Lu}}, \bibinfo {author} {\bibfnamefont
  {H.}~\bibnamefont {Zhang}}, \bibinfo {author} {\bibfnamefont
  {Y.}~\bibnamefont {Zhang}}, \bibinfo {author} {\bibfnamefont
  {K.}~\bibnamefont {Xia}},\ and\ \bibinfo {author} {\bibfnamefont
  {M.}~\bibnamefont {Xiao}},\ }\bibfield  {title} {\bibinfo {title} {On-chip
  chiral single-photon interface: Isolation and unidirectional emission},\
  }\href {https://doi.org/10.1103/PhysRevA.99.043833} {\bibfield  {journal}
  {\bibinfo  {journal} {Phys. Rev. A}\ }\textbf {\bibinfo {volume} {99}},\
  \bibinfo {pages} {043833} (\bibinfo {year} {2019})}\BibitemShut {NoStop}%
\bibitem [{\citenamefont {Ramos}\ \emph {et~al.}(2014)\citenamefont {Ramos},
  \citenamefont {Pichler}, \citenamefont {Daley},\ and\ \citenamefont
  {Zoller}}]{PhysRevLett.113.237203}%
  \BibitemOpen
  \bibfield  {author} {\bibinfo {author} {\bibfnamefont {T.}~\bibnamefont
  {Ramos}}, \bibinfo {author} {\bibfnamefont {H.}~\bibnamefont {Pichler}},
  \bibinfo {author} {\bibfnamefont {A.~J.}\ \bibnamefont {Daley}},\ and\
  \bibinfo {author} {\bibfnamefont {P.}~\bibnamefont {Zoller}},\ }\bibfield
  {title} {\bibinfo {title} {Quantum spin dimers from chiral dissipation in
  cold-atom chains},\ }\href {https://doi.org/10.1103/PhysRevLett.113.237203}
  {\bibfield  {journal} {\bibinfo  {journal} {Phys Rev Lett}\ }\textbf
  {\bibinfo {volume} {113}},\ \bibinfo {pages} {237203} (\bibinfo {year}
  {2014})}\BibitemShut {NoStop}%
\bibitem [{\citenamefont {S\"{o}llner}\ \emph {et~al.}(2015)\citenamefont
  {S\"{o}llner}, \citenamefont {Mahmoodian}, \citenamefont {Hansen},
  \citenamefont {Midolo}, \citenamefont {Javadi}, \citenamefont
  {Kir\v{s}ansk\.{e}}, \citenamefont {Pregnolato}, \citenamefont {El-Ella},
  \citenamefont {Lee}, \citenamefont {Song}, \citenamefont {Stobbe},\ and\
  \citenamefont {Lodahl}}]{nnano.2015.159}%
  \BibitemOpen
  \bibfield  {author} {\bibinfo {author} {\bibfnamefont {I.}~\bibnamefont
  {S\"{o}llner}}, \bibinfo {author} {\bibfnamefont {S.}~\bibnamefont
  {Mahmoodian}}, \bibinfo {author} {\bibfnamefont {S.~L.}\ \bibnamefont
  {Hansen}}, \bibinfo {author} {\bibfnamefont {L.}~\bibnamefont {Midolo}},
  \bibinfo {author} {\bibfnamefont {A.}~\bibnamefont {Javadi}}, \bibinfo
  {author} {\bibfnamefont {G.}~\bibnamefont {Kir\v{s}ansk\.{e}}}, \bibinfo
  {author} {\bibfnamefont {T.}~\bibnamefont {Pregnolato}}, \bibinfo {author}
  {\bibfnamefont {H.}~\bibnamefont {El-Ella}}, \bibinfo {author} {\bibfnamefont
  {E.~H.}\ \bibnamefont {Lee}}, \bibinfo {author} {\bibfnamefont {J.~D.}\
  \bibnamefont {Song}}, \bibinfo {author} {\bibfnamefont {S.}~\bibnamefont
  {Stobbe}},\ and\ \bibinfo {author} {\bibfnamefont {P.}~\bibnamefont
  {Lodahl}},\ }\bibfield  {title} {\bibinfo {title} {Deterministic
  photon-emitter coupling in chiral photonic circuits},\ }\href
  {https://doi.org/10.1038/nnano.2015.159} {\bibfield  {journal} {\bibinfo
  {journal} {Nat. Nanotechnol.}\ }\textbf {\bibinfo {volume} {10}},\ \bibinfo
  {pages} {775} (\bibinfo {year} {2015})}\BibitemShut {NoStop}%
\bibitem [{\citenamefont {Li}\ \emph {et~al.}(2021{\natexlab{a}})\citenamefont
  {Li}, \citenamefont {Gao},\ and\ \citenamefont {Xia}}]{OE.29.11.17613}%
  \BibitemOpen
  \bibfield  {author} {\bibinfo {author} {\bibfnamefont {T.}~\bibnamefont
  {Li}}, \bibinfo {author} {\bibfnamefont {Z.}~\bibnamefont {Gao}},\ and\
  \bibinfo {author} {\bibfnamefont {K.}~\bibnamefont {Xia}},\ }\bibfield
  {title} {\bibinfo {title} {Nonlinear-dissipation-induced nonreciprocal
  exceptional points},\ }\href {https://doi.org/10.1364/OE.426474} {\bibfield
  {journal} {\bibinfo  {journal} {Opt. Express}\ }\textbf {\bibinfo {volume}
  {29}},\ \bibinfo {pages} {17613} (\bibinfo {year}
  {2021}{\natexlab{a}})}\BibitemShut {NoStop}%
\bibitem [{\citenamefont {Tang}\ \emph
  {et~al.}(2021{\natexlab{c}})\citenamefont {Tang}, \citenamefont {Tang},
  \citenamefont {Wu}, \citenamefont {Wu}, \citenamefont {Sun}, \citenamefont
  {Zhang}, \citenamefont {Li}, \citenamefont {Lu}, \citenamefont {Xiao},\ and\
  \citenamefont {Xia}}]{PhysRevApplied.15.064020}%
  \BibitemOpen
  \bibfield  {author} {\bibinfo {author} {\bibfnamefont {J.}~\bibnamefont
  {Tang}}, \bibinfo {author} {\bibfnamefont {L.}~\bibnamefont {Tang}}, \bibinfo
  {author} {\bibfnamefont {H.}~\bibnamefont {Wu}}, \bibinfo {author}
  {\bibfnamefont {Y.}~\bibnamefont {Wu}}, \bibinfo {author} {\bibfnamefont
  {H.}~\bibnamefont {Sun}}, \bibinfo {author} {\bibfnamefont {H.}~\bibnamefont
  {Zhang}}, \bibinfo {author} {\bibfnamefont {T.}~\bibnamefont {Li}}, \bibinfo
  {author} {\bibfnamefont {Y.}~\bibnamefont {Lu}}, \bibinfo {author}
  {\bibfnamefont {M.}~\bibnamefont {Xiao}},\ and\ \bibinfo {author}
  {\bibfnamefont {K.}~\bibnamefont {Xia}},\ }\bibfield  {title} {\bibinfo
  {title} {Towards on-demand heralded single-photon sources via photon
  blockade},\ }\href {https://doi.org/10.1103/PhysRevApplied.15.064020}
  {\bibfield  {journal} {\bibinfo  {journal} {Phys. Rev. Appl.}\ }\textbf
  {\bibinfo {volume} {15}},\ \bibinfo {pages} {064020} (\bibinfo {year}
  {2021}{\natexlab{c}})}\BibitemShut {NoStop}%
\bibitem [{\citenamefont {Roy}\ \emph {et~al.}(2017)\citenamefont {Roy},
  \citenamefont {Wilson},\ and\ \citenamefont
  {Firstenberg}}]{RevModPhys.89.021001}%
  \BibitemOpen
  \bibfield  {author} {\bibinfo {author} {\bibfnamefont {D.}~\bibnamefont
  {Roy}}, \bibinfo {author} {\bibfnamefont {C.~M.}\ \bibnamefont {Wilson}},\
  and\ \bibinfo {author} {\bibfnamefont {O.}~\bibnamefont {Firstenberg}},\
  }\bibfield  {title} {\bibinfo {title} {Colloquium: Strongly interacting
  photons in one-dimensional continuum},\ }\href
  {https://doi.org/10.1103/RevModPhys.89.021001} {\bibfield  {journal}
  {\bibinfo  {journal} {Rev. Mod. Phys.}\ }\textbf {\bibinfo {volume} {89}},\
  \bibinfo {pages} {021001} (\bibinfo {year} {2017})}\BibitemShut {NoStop}%
\bibitem [{\citenamefont {Chang}\ \emph {et~al.}(2018)\citenamefont {Chang},
  \citenamefont {Douglas}, \citenamefont {Gonz\'alez-Tudela}, \citenamefont
  {Hung},\ and\ \citenamefont {Kimble}}]{RevModPhys.90.031002}%
  \BibitemOpen
  \bibfield  {author} {\bibinfo {author} {\bibfnamefont {D.~E.}\ \bibnamefont
  {Chang}}, \bibinfo {author} {\bibfnamefont {J.~S.}\ \bibnamefont {Douglas}},
  \bibinfo {author} {\bibfnamefont {A.}~\bibnamefont {Gonz\'alez-Tudela}},
  \bibinfo {author} {\bibfnamefont {C.-L.}\ \bibnamefont {Hung}},\ and\
  \bibinfo {author} {\bibfnamefont {H.~J.}\ \bibnamefont {Kimble}},\ }\bibfield
   {title} {\bibinfo {title} {Colloquium: Quantum matter built from nanoscopic
  lattices of atoms and photons},\ }\href
  {https://doi.org/10.1103/RevModPhys.90.031002} {\bibfield  {journal}
  {\bibinfo  {journal} {Rev. Mod. Phys.}\ }\textbf {\bibinfo {volume} {90}},\
  \bibinfo {pages} {031002} (\bibinfo {year} {2018})}\BibitemShut {NoStop}%
\bibitem [{\citenamefont {Yariv}\ \emph {et~al.}(1999)\citenamefont {Yariv},
  \citenamefont {Xu}, \citenamefont {Lee},\ and\ \citenamefont
  {Scherer}}]{OL.24.000711}%
  \BibitemOpen
  \bibfield  {author} {\bibinfo {author} {\bibfnamefont {A.}~\bibnamefont
  {Yariv}}, \bibinfo {author} {\bibfnamefont {Y.}~\bibnamefont {Xu}}, \bibinfo
  {author} {\bibfnamefont {R.~K.}\ \bibnamefont {Lee}},\ and\ \bibinfo {author}
  {\bibfnamefont {A.}~\bibnamefont {Scherer}},\ }\bibfield  {title} {\bibinfo
  {title} {Coupled-resonator optical waveguide: a proposal and analysis},\
  }\href {https://doi.org/10.1364/OL.24.000711} {\bibfield  {journal} {\bibinfo
   {journal} {Opt. Lett.}\ }\textbf {\bibinfo {volume} {24}},\ \bibinfo {pages}
  {711} (\bibinfo {year} {1999})}\BibitemShut {NoStop}%
\bibitem [{\citenamefont {Poon}\ \emph
  {et~al.}(2004{\natexlab{a}})\citenamefont {Poon}, \citenamefont {Scheuer},
  \citenamefont {Mookherjea}, \citenamefont {Paloczi}, \citenamefont {Huang},\
  and\ \citenamefont {Yariv}}]{OE.12.000090}%
  \BibitemOpen
  \bibfield  {author} {\bibinfo {author} {\bibfnamefont {J.~K.~S.}\
  \bibnamefont {Poon}}, \bibinfo {author} {\bibfnamefont {J.}~\bibnamefont
  {Scheuer}}, \bibinfo {author} {\bibfnamefont {S.}~\bibnamefont {Mookherjea}},
  \bibinfo {author} {\bibfnamefont {G.~T.}\ \bibnamefont {Paloczi}}, \bibinfo
  {author} {\bibfnamefont {Y.}~\bibnamefont {Huang}},\ and\ \bibinfo {author}
  {\bibfnamefont {A.}~\bibnamefont {Yariv}},\ }\bibfield  {title} {\bibinfo
  {title} {Matrix analysis of microring coupled-resonator optical waveguides},\
  }\href {https://doi.org/10.1364/OPEX.12.000090} {\bibfield  {journal}
  {\bibinfo  {journal} {Opt. Express}\ }\textbf {\bibinfo {volume} {12}},\
  \bibinfo {pages} {90} (\bibinfo {year} {2004}{\natexlab{a}})}\BibitemShut
  {NoStop}%
\bibitem [{\citenamefont {Perczel}\ \emph {et~al.}(2020)\citenamefont
  {Perczel}, \citenamefont {Borregaard}, \citenamefont {Chang}, \citenamefont
  {Yelin},\ and\ \citenamefont {Lukin}}]{PhysRevLett.124.083603}%
  \BibitemOpen
  \bibfield  {author} {\bibinfo {author} {\bibfnamefont {J.}~\bibnamefont
  {Perczel}}, \bibinfo {author} {\bibfnamefont {J.}~\bibnamefont {Borregaard}},
  \bibinfo {author} {\bibfnamefont {D.~E.}\ \bibnamefont {Chang}}, \bibinfo
  {author} {\bibfnamefont {S.~F.}\ \bibnamefont {Yelin}},\ and\ \bibinfo
  {author} {\bibfnamefont {M.~D.}\ \bibnamefont {Lukin}},\ }\bibfield  {title}
  {\bibinfo {title} {Topological quantum optics using atomlike emitter arrays
  coupled to photonic crystals},\ }\href
  {https://doi.org/10.1103/PhysRevLett.124.083603} {\bibfield  {journal}
  {\bibinfo  {journal} {Phys. Rev. Lett.}\ }\textbf {\bibinfo {volume} {124}},\
  \bibinfo {pages} {083603} (\bibinfo {year} {2020})}\BibitemShut {NoStop}%
\bibitem [{\citenamefont {Ao}\ \emph {et~al.}(2020)\citenamefont {Ao},
  \citenamefont {Hu}, \citenamefont {You}, \citenamefont {Lu}, \citenamefont
  {Fu}, \citenamefont {Wang},\ and\ \citenamefont
  {Gong}}]{PhysRevLett.125.013902}%
  \BibitemOpen
  \bibfield  {author} {\bibinfo {author} {\bibfnamefont {Y.}~\bibnamefont
  {Ao}}, \bibinfo {author} {\bibfnamefont {X.}~\bibnamefont {Hu}}, \bibinfo
  {author} {\bibfnamefont {Y.}~\bibnamefont {You}}, \bibinfo {author}
  {\bibfnamefont {C.}~\bibnamefont {Lu}}, \bibinfo {author} {\bibfnamefont
  {Y.}~\bibnamefont {Fu}}, \bibinfo {author} {\bibfnamefont {X.}~\bibnamefont
  {Wang}},\ and\ \bibinfo {author} {\bibfnamefont {Q.}~\bibnamefont {Gong}},\
  }\bibfield  {title} {\bibinfo {title} {Topological phase transition in the
  non-{H}ermitian coupled resonator array},\ }\href
  {https://doi.org/10.1103/PhysRevLett.125.013902} {\bibfield  {journal}
  {\bibinfo  {journal} {Phys. Rev. Lett.}\ }\textbf {\bibinfo {volume} {125}},\
  \bibinfo {pages} {013902} (\bibinfo {year} {2020})}\BibitemShut {NoStop}%
\bibitem [{\citenamefont {Li}\ \emph {et~al.}(2021{\natexlab{b}})\citenamefont
  {Li}, \citenamefont {Gao}, \citenamefont {Zhu}, \citenamefont {Xu},\ and\
  \citenamefont {Yang}}]{APL.119.14.141108}%
  \BibitemOpen
  \bibfield  {author} {\bibinfo {author} {\bibfnamefont {J.}~\bibnamefont
  {Li}}, \bibinfo {author} {\bibfnamefont {B.}~\bibnamefont {Gao}}, \bibinfo
  {author} {\bibfnamefont {C.}~\bibnamefont {Zhu}}, \bibinfo {author}
  {\bibfnamefont {J.}~\bibnamefont {Xu}},\ and\ \bibinfo {author}
  {\bibfnamefont {Y.}~\bibnamefont {Yang}},\ }\bibfield  {title} {\bibinfo
  {title} {Nonreciprocal photonic composited {Su-Schrieffer-Heeger} chain},\
  }\href {https://doi.org/10.1063/5.0063247} {\bibfield  {journal} {\bibinfo
  {journal} {Appl. Phys. Lett.}\ }\textbf {\bibinfo {volume} {119}},\ \bibinfo
  {pages} {141108} (\bibinfo {year} {2021}{\natexlab{b}})}\BibitemShut
  {NoStop}%
\bibitem [{\citenamefont {Fang}\ \emph {et~al.}(2011)\citenamefont {Fang},
  \citenamefont {Yu},\ and\ \citenamefont {Fan}}]{PhysRevB.84.075477}%
  \BibitemOpen
  \bibfield  {author} {\bibinfo {author} {\bibfnamefont {K.}~\bibnamefont
  {Fang}}, \bibinfo {author} {\bibfnamefont {Z.}~\bibnamefont {Yu}},\ and\
  \bibinfo {author} {\bibfnamefont {S.}~\bibnamefont {Fan}},\ }\bibfield
  {title} {\bibinfo {title} {Microscopic theory of photonic one-way edge
  mode},\ }\href {https://doi.org/10.1103/PhysRevB.84.075477} {\bibfield
  {journal} {\bibinfo  {journal} {Phys. Rev. B}\ }\textbf {\bibinfo {volume}
  {84}},\ \bibinfo {pages} {075477} (\bibinfo {year} {2011})}\BibitemShut
  {NoStop}%
\bibitem [{\citenamefont {Lu}\ \emph {et~al.}(2014)\citenamefont {Lu},
  \citenamefont {Joannopoulos},\ and\ \citenamefont
  {Solja\v{c}i\'{o}}}]{nphoton.2014.248}%
  \BibitemOpen
  \bibfield  {author} {\bibinfo {author} {\bibfnamefont {L.}~\bibnamefont
  {Lu}}, \bibinfo {author} {\bibfnamefont {J.~D.}\ \bibnamefont
  {Joannopoulos}},\ and\ \bibinfo {author} {\bibfnamefont {M.}~\bibnamefont
  {Solja\v{c}i\'{o}}},\ }\bibfield  {title} {\bibinfo {title} {Topological
  photonics},\ }\href {https://doi.org/10.1038/nphoton.2014.248} {\bibfield
  {journal} {\bibinfo  {journal} {Nat. Photonics}\ }\textbf {\bibinfo {volume}
  {8}},\ \bibinfo {pages} {821} (\bibinfo {year} {2014})}\BibitemShut {NoStop}%
\bibitem [{\citenamefont {Khanikaev}\ and\ \citenamefont
  {Shvets}(2017)}]{nphotons.11.12}%
  \BibitemOpen
  \bibfield  {author} {\bibinfo {author} {\bibfnamefont {A.~B.}\ \bibnamefont
  {Khanikaev}}\ and\ \bibinfo {author} {\bibfnamefont {G.}~\bibnamefont
  {Shvets}},\ }\bibfield  {title} {\bibinfo {title} {Two-dimensional
  topological photonics},\ }\href {https://doi.org/10.1038/s41566-017-0048-5}
  {\bibfield  {journal} {\bibinfo  {journal} {Nat. Photonics}\ }\textbf
  {\bibinfo {volume} {11}},\ \bibinfo {pages} {763} (\bibinfo {year}
  {2017})}\BibitemShut {NoStop}%
\bibitem [{\citenamefont {Ozawa}\ \emph {et~al.}(2019)\citenamefont {Ozawa},
  \citenamefont {Price}, \citenamefont {Amo}, \citenamefont {Goldman},
  \citenamefont {Hafezi}, \citenamefont {Lu}, \citenamefont {Rechtsman},
  \citenamefont {Schuster}, \citenamefont {Simon}, \citenamefont {Zilberberg},\
  and\ \citenamefont {Carusotto}}]{RevModPhys.91.015006}%
  \BibitemOpen
  \bibfield  {author} {\bibinfo {author} {\bibfnamefont {T.}~\bibnamefont
  {Ozawa}}, \bibinfo {author} {\bibfnamefont {H.~M.}\ \bibnamefont {Price}},
  \bibinfo {author} {\bibfnamefont {A.}~\bibnamefont {Amo}}, \bibinfo {author}
  {\bibfnamefont {N.}~\bibnamefont {Goldman}}, \bibinfo {author} {\bibfnamefont
  {M.}~\bibnamefont {Hafezi}}, \bibinfo {author} {\bibfnamefont
  {L.}~\bibnamefont {Lu}}, \bibinfo {author} {\bibfnamefont {M.~C.}\
  \bibnamefont {Rechtsman}}, \bibinfo {author} {\bibfnamefont {D.}~\bibnamefont
  {Schuster}}, \bibinfo {author} {\bibfnamefont {J.}~\bibnamefont {Simon}},
  \bibinfo {author} {\bibfnamefont {O.}~\bibnamefont {Zilberberg}},\ and\
  \bibinfo {author} {\bibfnamefont {I.}~\bibnamefont {Carusotto}},\ }\bibfield
  {title} {\bibinfo {title} {Topological photonics},\ }\href
  {https://doi.org/10.1103/RevModPhys.91.015006} {\bibfield  {journal}
  {\bibinfo  {journal} {Rev. Mod. Phys.}\ }\textbf {\bibinfo {volume} {91}},\
  \bibinfo {pages} {015006} (\bibinfo {year} {2019})}\BibitemShut {NoStop}%
\bibitem [{\citenamefont {Xi}\ \emph {et~al.}(2021)\citenamefont {Xi},
  \citenamefont {Ma}, \citenamefont {Wan}, \citenamefont {Dong},\ and\
  \citenamefont {Sun}}]{sciadv.abe1398}%
  \BibitemOpen
  \bibfield  {author} {\bibinfo {author} {\bibfnamefont {X.}~\bibnamefont
  {Xi}}, \bibinfo {author} {\bibfnamefont {J.}~\bibnamefont {Ma}}, \bibinfo
  {author} {\bibfnamefont {S.}~\bibnamefont {Wan}}, \bibinfo {author}
  {\bibfnamefont {C.-H.}\ \bibnamefont {Dong}},\ and\ \bibinfo {author}
  {\bibfnamefont {X.}~\bibnamefont {Sun}},\ }\bibfield  {title} {\bibinfo
  {title} {Observation of chiral edge states in gapped nanomechanical
  graphene},\ }\href {https://doi.org/10.1126/sciadv.abe1398} {\bibfield
  {journal} {\bibinfo  {journal} {Sci. Adv.}\ }\textbf {\bibinfo {volume}
  {7}},\ \bibinfo {pages} {eabe1398} (\bibinfo {year} {2021})}\BibitemShut
  {NoStop}%
\bibitem [{\citenamefont {Bliokh}\ and\ \citenamefont
  {Nori}(2011)}]{PhysRevA.83.021803}%
  \BibitemOpen
  \bibfield  {author} {\bibinfo {author} {\bibfnamefont {K.~Y.}\ \bibnamefont
  {Bliokh}}\ and\ \bibinfo {author} {\bibfnamefont {F.}~\bibnamefont {Nori}},\
  }\bibfield  {title} {\bibinfo {title} {Characterizing optical chirality},\
  }\href {https://doi.org/10.1103/PhysRevA.83.021803} {\bibfield  {journal}
  {\bibinfo  {journal} {Phys. Rev. A}\ }\textbf {\bibinfo {volume} {83}},\
  \bibinfo {pages} {021803} (\bibinfo {year} {2011})}\BibitemShut {NoStop}%
\bibitem [{\citenamefont {Bliokh}\ \emph {et~al.}(2014)\citenamefont {Bliokh},
  \citenamefont {Bekshaev},\ and\ \citenamefont {Nori}}]{ncomms.5.1}%
  \BibitemOpen
  \bibfield  {author} {\bibinfo {author} {\bibfnamefont {K.~Y.}\ \bibnamefont
  {Bliokh}}, \bibinfo {author} {\bibfnamefont {A.~Y.}\ \bibnamefont
  {Bekshaev}},\ and\ \bibinfo {author} {\bibfnamefont {F.}~\bibnamefont
  {Nori}},\ }\bibfield  {title} {\bibinfo {title} {Extraordinary momentum and
  spin in evanescent waves},\ }\href {https://doi.org/10.1038/ncomms4300}
  {\bibfield  {journal} {\bibinfo  {journal} {Nat. Commun.}\ }\textbf {\bibinfo
  {volume} {5}},\ \bibinfo {pages} {3300} (\bibinfo {year} {2014})}\BibitemShut
  {NoStop}%
\bibitem [{\citenamefont {Bliokh}\ \emph {et~al.}(2015)\citenamefont {Bliokh},
  \citenamefont {Smirnova},\ and\ \citenamefont {Nori}}]{science.348.1148}%
  \BibitemOpen
  \bibfield  {author} {\bibinfo {author} {\bibfnamefont {K.~Y.}\ \bibnamefont
  {Bliokh}}, \bibinfo {author} {\bibfnamefont {D.}~\bibnamefont {Smirnova}},\
  and\ \bibinfo {author} {\bibfnamefont {F.}~\bibnamefont {Nori}},\ }\bibfield
  {title} {\bibinfo {title} {Quantum spin {Hall} effect of light},\ }\href
  {https://doi.org/10.1126/science.aaa9519} {\bibfield  {journal} {\bibinfo
  {journal} {Science}\ }\textbf {\bibinfo {volume} {348}},\ \bibinfo {pages}
  {1448} (\bibinfo {year} {2015})}\BibitemShut {NoStop}%
\bibitem [{\citenamefont {Alpeggiani}\ \emph {et~al.}(2018)\citenamefont
  {Alpeggiani}, \citenamefont {Bliokh}, \citenamefont {Nori},\ and\
  \citenamefont {Kuipers}}]{PhysRevLett.120.243605}%
  \BibitemOpen
  \bibfield  {author} {\bibinfo {author} {\bibfnamefont {F.}~\bibnamefont
  {Alpeggiani}}, \bibinfo {author} {\bibfnamefont {K.~Y.}\ \bibnamefont
  {Bliokh}}, \bibinfo {author} {\bibfnamefont {F.}~\bibnamefont {Nori}},\ and\
  \bibinfo {author} {\bibfnamefont {L.}~\bibnamefont {Kuipers}},\ }\bibfield
  {title} {\bibinfo {title} {Electromagnetic helicity in complex media},\
  }\href {https://doi.org/10.1103/PhysRevLett.120.243605} {\bibfield  {journal}
  {\bibinfo  {journal} {Phys. Rev. Lett.}\ }\textbf {\bibinfo {volume} {120}},\
  \bibinfo {pages} {243605} (\bibinfo {year} {2018})}\BibitemShut {NoStop}%
\bibitem [{\citenamefont {Poon}\ \emph
  {et~al.}(2004{\natexlab{b}})\citenamefont {Poon}, \citenamefont {Scheuer},
  \citenamefont {Xu},\ and\ \citenamefont {Yariv}}]{JOSAB.21.001665}%
  \BibitemOpen
  \bibfield  {author} {\bibinfo {author} {\bibfnamefont {J.~K.~S.}\
  \bibnamefont {Poon}}, \bibinfo {author} {\bibfnamefont {J.}~\bibnamefont
  {Scheuer}}, \bibinfo {author} {\bibfnamefont {Y.}~\bibnamefont {Xu}},\ and\
  \bibinfo {author} {\bibfnamefont {A.}~\bibnamefont {Yariv}},\ }\bibfield
  {title} {\bibinfo {title} {Designing coupled-resonator optical waveguide
  delay lines},\ }\href {https://doi.org/10.1364/JOSAB.21.001665} {\bibfield
  {journal} {\bibinfo  {journal} {J. Opt. Soc. Am. B}\ }\textbf {\bibinfo
  {volume} {21}},\ \bibinfo {pages} {1665} (\bibinfo {year}
  {2004}{\natexlab{b}})}\BibitemShut {NoStop}%
\bibitem [{\citenamefont {Thompson}\ \emph {et~al.}(2013)\citenamefont
  {Thompson}, \citenamefont {Tiecke}, \citenamefont {de~Leon}, \citenamefont
  {Feist}, \citenamefont {Akimov}, \citenamefont {Gullans}, \citenamefont
  {Zibrov}, \citenamefont {Vuleti\'{c}},\ and\ \citenamefont
  {Lukin}}]{science.1237125}%
  \BibitemOpen
  \bibfield  {author} {\bibinfo {author} {\bibfnamefont {J.~D.}\ \bibnamefont
  {Thompson}}, \bibinfo {author} {\bibfnamefont {T.~G.}\ \bibnamefont
  {Tiecke}}, \bibinfo {author} {\bibfnamefont {N.~P.}\ \bibnamefont {de~Leon}},
  \bibinfo {author} {\bibfnamefont {J.}~\bibnamefont {Feist}}, \bibinfo
  {author} {\bibfnamefont {A.~V.}\ \bibnamefont {Akimov}}, \bibinfo {author}
  {\bibfnamefont {M.}~\bibnamefont {Gullans}}, \bibinfo {author} {\bibfnamefont
  {A.~S.}\ \bibnamefont {Zibrov}}, \bibinfo {author} {\bibfnamefont
  {V.}~\bibnamefont {Vuleti\'{c}}},\ and\ \bibinfo {author} {\bibfnamefont
  {M.~D.}\ \bibnamefont {Lukin}},\ }\bibfield  {title} {\bibinfo {title}
  {Coupling a single trapped atom to a nanoscale optical cavity},\ }\href
  {https://doi.org/10.1126/science.1237125} {\bibfield  {journal} {\bibinfo
  {journal} {Science}\ }\textbf {\bibinfo {volume} {340}},\ \bibinfo {pages}
  {1202} (\bibinfo {year} {2013})}\BibitemShut {NoStop}%
\bibitem [{\citenamefont {Barik}\ \emph {et~al.}(2018)\citenamefont {Barik},
  \citenamefont {Karasahin}, \citenamefont {Flower}, \citenamefont {Cai},
  \citenamefont {Miyake}, \citenamefont {DeGottardi}, \citenamefont {Hafezi},\
  and\ \citenamefont {Waks}}]{science.aaq0327}%
  \BibitemOpen
  \bibfield  {author} {\bibinfo {author} {\bibfnamefont {S.}~\bibnamefont
  {Barik}}, \bibinfo {author} {\bibfnamefont {A.}~\bibnamefont {Karasahin}},
  \bibinfo {author} {\bibfnamefont {C.}~\bibnamefont {Flower}}, \bibinfo
  {author} {\bibfnamefont {T.}~\bibnamefont {Cai}}, \bibinfo {author}
  {\bibfnamefont {H.}~\bibnamefont {Miyake}}, \bibinfo {author} {\bibfnamefont
  {W.}~\bibnamefont {DeGottardi}}, \bibinfo {author} {\bibfnamefont
  {M.}~\bibnamefont {Hafezi}},\ and\ \bibinfo {author} {\bibfnamefont
  {E.}~\bibnamefont {Waks}},\ }\bibfield  {title} {\bibinfo {title} {A
  topological quantum optics interface},\ }\href
  {https://doi.org/10.1126/science.aaq0327} {\bibfield  {journal} {\bibinfo
  {journal} {Science}\ }\textbf {\bibinfo {volume} {359}},\ \bibinfo {pages}
  {666} (\bibinfo {year} {2018})}\BibitemShut {NoStop}%
\bibitem [{\citenamefont {Yang}\ \emph {et~al.}(2020)\citenamefont {Yang},
  \citenamefont {Qian}, \citenamefont {Xie}, \citenamefont {Peng},
  \citenamefont {Wu}, \citenamefont {Song}, \citenamefont {Sun}, \citenamefont
  {Dang}, \citenamefont {Yu}, \citenamefont {Shi}, \citenamefont {He},
  \citenamefont {Steer}, \citenamefont {Thayne}, \citenamefont {Li},
  \citenamefont {Bo}, \citenamefont {Xiao}, \citenamefont {Zuo}, \citenamefont
  {Jin}, \citenamefont {Gu},\ and\ \citenamefont {Xu}}]{s41377-020-0244-9}%
  \BibitemOpen
  \bibfield  {author} {\bibinfo {author} {\bibfnamefont {J.}~\bibnamefont
  {Yang}}, \bibinfo {author} {\bibfnamefont {C.}~\bibnamefont {Qian}}, \bibinfo
  {author} {\bibfnamefont {X.}~\bibnamefont {Xie}}, \bibinfo {author}
  {\bibfnamefont {K.}~\bibnamefont {Peng}}, \bibinfo {author} {\bibfnamefont
  {S.}~\bibnamefont {Wu}}, \bibinfo {author} {\bibfnamefont {F.}~\bibnamefont
  {Song}}, \bibinfo {author} {\bibfnamefont {S.}~\bibnamefont {Sun}}, \bibinfo
  {author} {\bibfnamefont {J.}~\bibnamefont {Dang}}, \bibinfo {author}
  {\bibfnamefont {Y.}~\bibnamefont {Yu}}, \bibinfo {author} {\bibfnamefont
  {S.}~\bibnamefont {Shi}}, \bibinfo {author} {\bibfnamefont {J.}~\bibnamefont
  {He}}, \bibinfo {author} {\bibfnamefont {M.~J.}\ \bibnamefont {Steer}},
  \bibinfo {author} {\bibfnamefont {I.~G.}\ \bibnamefont {Thayne}}, \bibinfo
  {author} {\bibfnamefont {B.-B.}\ \bibnamefont {Li}}, \bibinfo {author}
  {\bibfnamefont {F.}~\bibnamefont {Bo}}, \bibinfo {author} {\bibfnamefont
  {Y.-F.}\ \bibnamefont {Xiao}}, \bibinfo {author} {\bibfnamefont
  {Z.}~\bibnamefont {Zuo}}, \bibinfo {author} {\bibfnamefont {K.}~\bibnamefont
  {Jin}}, \bibinfo {author} {\bibfnamefont {C.}~\bibnamefont {Gu}},\ and\
  \bibinfo {author} {\bibfnamefont {X.}~\bibnamefont {Xu}},\ }\bibfield
  {title} {\bibinfo {title} {Diabolical points in coupled active cavities with
  quantum emitters},\ }\href {https://doi.org/10.1038/s41377-020-0244-9}
  {\bibfield  {journal} {\bibinfo  {journal} {Light-Sci. Appl.}\ }\textbf
  {\bibinfo {volume} {9}},\ \bibinfo {pages} {6} (\bibinfo {year}
  {2020})}\BibitemShut {NoStop}%
\bibitem [{\citenamefont {Branny}\ \emph {et~al.}(2017)\citenamefont {Branny},
  \citenamefont {Kumar}, \citenamefont {Proux},\ and\ \citenamefont
  {Gerardot}}]{ncomms15053}%
  \BibitemOpen
  \bibfield  {author} {\bibinfo {author} {\bibfnamefont {A.}~\bibnamefont
  {Branny}}, \bibinfo {author} {\bibfnamefont {S.}~\bibnamefont {Kumar}},
  \bibinfo {author} {\bibfnamefont {R.}~\bibnamefont {Proux}},\ and\ \bibinfo
  {author} {\bibfnamefont {B.~D.}\ \bibnamefont {Gerardot}},\ }\bibfield
  {title} {\bibinfo {title} {Deterministic strain-induced arrays of quantum
  emitters in a two-dimensional semiconductor},\ }\href
  {https://doi.org/10.1038/ncomms15053} {\bibfield  {journal} {\bibinfo
  {journal} {Nat. Commun.}\ }\textbf {\bibinfo {volume} {8}},\ \bibinfo {pages}
  {15053} (\bibinfo {year} {2017})}\BibitemShut {NoStop}%
\bibitem [{\citenamefont {Palacios-Berraquero}\ \emph
  {et~al.}(2017)\citenamefont {Palacios-Berraquero}, \citenamefont {Kara},
  \citenamefont {Montblanch}, \citenamefont {Barbone}, \citenamefont
  {Latawiec}, \citenamefont {Yoon}, \citenamefont {Ott}, \citenamefont
  {Loncar}, \citenamefont {Ferrari},\ and\ \citenamefont
  {Atat\"{u}re}}]{ncomms15093}%
  \BibitemOpen
  \bibfield  {author} {\bibinfo {author} {\bibfnamefont {C.}~\bibnamefont
  {Palacios-Berraquero}}, \bibinfo {author} {\bibfnamefont {D.~M.}\
  \bibnamefont {Kara}}, \bibinfo {author} {\bibfnamefont {A.~R.~P.}\
  \bibnamefont {Montblanch}}, \bibinfo {author} {\bibfnamefont
  {M.}~\bibnamefont {Barbone}}, \bibinfo {author} {\bibfnamefont
  {P.}~\bibnamefont {Latawiec}}, \bibinfo {author} {\bibfnamefont
  {D.}~\bibnamefont {Yoon}}, \bibinfo {author} {\bibfnamefont {A.~K.}\
  \bibnamefont {Ott}}, \bibinfo {author} {\bibfnamefont {M.}~\bibnamefont
  {Loncar}}, \bibinfo {author} {\bibfnamefont {A.~C.}\ \bibnamefont
  {Ferrari}},\ and\ \bibinfo {author} {\bibfnamefont {M.}~\bibnamefont
  {Atat\"{u}re}},\ }\bibfield  {title} {\bibinfo {title} {Large-scale
  quantum-emitter arrays in atomically thin semiconductors},\ }\href
  {https://doi.org/10.1038/ncomms15093} {\bibfield  {journal} {\bibinfo
  {journal} {Nat. Commun.}\ }\textbf {\bibinfo {volume} {8}},\ \bibinfo {pages}
  {15093} (\bibinfo {year} {2017})}\BibitemShut {NoStop}%
\bibitem [{\citenamefont {Su}\ \emph {et~al.}(1979)\citenamefont {Su},
  \citenamefont {Schrieffer},\ and\ \citenamefont
  {Heeger}}]{PhysRevLett.42.1698}%
  \BibitemOpen
  \bibfield  {author} {\bibinfo {author} {\bibfnamefont {W.~P.}\ \bibnamefont
  {Su}}, \bibinfo {author} {\bibfnamefont {J.~R.}\ \bibnamefont {Schrieffer}},\
  and\ \bibinfo {author} {\bibfnamefont {A.~J.}\ \bibnamefont {Heeger}},\
  }\bibfield  {title} {\bibinfo {title} {Solitons in polyacetylene},\ }\href
  {https://doi.org/10.1103/PhysRevLett.42.1698} {\bibfield  {journal} {\bibinfo
   {journal} {Phys. Rev. Lett.}\ }\textbf {\bibinfo {volume} {42}},\ \bibinfo
  {pages} {1698} (\bibinfo {year} {1979})}\BibitemShut {NoStop}%
\bibitem [{\citenamefont {Jiao}\ \emph {et~al.}(2021)\citenamefont {Jiao},
  \citenamefont {Longhi}, \citenamefont {Wang}, \citenamefont {Gao},
  \citenamefont {Zhou}, \citenamefont {Wang}, \citenamefont {Fu}, \citenamefont
  {Wang}, \citenamefont {Ren}, \citenamefont {Qiao},\ and\ \citenamefont
  {Jin}}]{PhysRevLett.127.147401}%
  \BibitemOpen
  \bibfield  {author} {\bibinfo {author} {\bibfnamefont {Z.-Q.}\ \bibnamefont
  {Jiao}}, \bibinfo {author} {\bibfnamefont {S.}~\bibnamefont {Longhi}},
  \bibinfo {author} {\bibfnamefont {X.-W.}\ \bibnamefont {Wang}}, \bibinfo
  {author} {\bibfnamefont {J.}~\bibnamefont {Gao}}, \bibinfo {author}
  {\bibfnamefont {W.-H.}\ \bibnamefont {Zhou}}, \bibinfo {author}
  {\bibfnamefont {Y.}~\bibnamefont {Wang}}, \bibinfo {author} {\bibfnamefont
  {Y.-X.}\ \bibnamefont {Fu}}, \bibinfo {author} {\bibfnamefont
  {L.}~\bibnamefont {Wang}}, \bibinfo {author} {\bibfnamefont {R.-J.}\
  \bibnamefont {Ren}}, \bibinfo {author} {\bibfnamefont {L.-F.}\ \bibnamefont
  {Qiao}},\ and\ \bibinfo {author} {\bibfnamefont {X.-M.}\ \bibnamefont
  {Jin}},\ }\bibfield  {title} {\bibinfo {title} {Experimentally detecting
  quantized {Zak} phases without chiral symmetry in photonic lattices},\ }\href
  {https://doi.org/10.1103/PhysRevLett.127.147401} {\bibfield  {journal}
  {\bibinfo  {journal} {Phys. Rev. Lett.}\ }\textbf {\bibinfo {volume} {127}},\
  \bibinfo {pages} {147401} (\bibinfo {year} {2021})}\BibitemShut {NoStop}%
\bibitem [{Sup()}]{SupplMat}%
  \BibitemOpen
  \href@noop {} {\bibinfo  {journal} {See Supplemental Material for the
  detailed derivation}\ }\BibitemShut {NoStop}%
\bibitem [{\citenamefont {Tang}\ \emph
  {et~al.}(2021{\natexlab{d}})\citenamefont {Tang}, \citenamefont {Tang},\ and\
  \citenamefont {Xia}}]{tang2021}%
  \BibitemOpen
\bibfield  {journal} {  }\bibfield  {author} {\bibinfo {author} {\bibfnamefont
  {J.}~\bibnamefont {Tang}}, \bibinfo {author} {\bibfnamefont {L.}~\bibnamefont
  {Tang}},\ and\ \bibinfo {author} {\bibfnamefont {K.}~\bibnamefont {Xia}},\
  }\bibfield  {title} {\bibinfo {title} {Single-photon transport in a
  whispering-gallery mode microresonator directionally coupled with a two-level
  quantum emitter},\ }\href {https://arxiv.org/abs/2110.09375} {\bibfield
  {journal} {\bibinfo  {journal} {arXiv preprint arXiv:2110.09375}\ } (\bibinfo
  {year} {2021}{\natexlab{d}})}\BibitemShut {NoStop}%
\bibitem [{\citenamefont {Asb{\'o}th}\ \emph {et~al.}(2016)\citenamefont
  {Asb{\'o}th}, \citenamefont {Oroszl{\'a}ny},\ and\ \citenamefont
  {P{\'a}lyi}}]{asboth2016short}%
  \BibitemOpen
  \bibfield  {author} {\bibinfo {author} {\bibfnamefont {J.~K.}\ \bibnamefont
  {Asb{\'o}th}}, \bibinfo {author} {\bibfnamefont {L.}~\bibnamefont
  {Oroszl{\'a}ny}},\ and\ \bibinfo {author} {\bibfnamefont {A.}~\bibnamefont
  {P{\'a}lyi}},\ }\bibfield  {title} {\bibinfo {title} {A short course on
  topological insulators},\ }\href@noop {} {\bibfield  {journal} {\bibinfo
  {journal} {Lecture notes in physics}\ }\textbf {\bibinfo {volume} {919}},\
  \bibinfo {pages} {166} (\bibinfo {year} {2016})}\BibitemShut {NoStop}%
\bibitem [{\citenamefont {Yanik}\ \emph {et~al.}(2004)\citenamefont {Yanik},
  \citenamefont {Suh}, \citenamefont {Wang},\ and\ \citenamefont
  {Fan}}]{PhysRevLett.93.233903}%
  \BibitemOpen
  \bibfield  {author} {\bibinfo {author} {\bibfnamefont {M.~F.}\ \bibnamefont
  {Yanik}}, \bibinfo {author} {\bibfnamefont {W.}~\bibnamefont {Suh}}, \bibinfo
  {author} {\bibfnamefont {Z.}~\bibnamefont {Wang}},\ and\ \bibinfo {author}
  {\bibfnamefont {S.}~\bibnamefont {Fan}},\ }\bibfield  {title} {\bibinfo
  {title} {Stopping light in a waveguide with an all-optical analog of
  electromagnetically induced transparency},\ }\href
  {https://doi.org/10.1103/PhysRevLett.93.233903} {\bibfield  {journal}
  {\bibinfo  {journal} {Phys. Rev. Lett.}\ }\textbf {\bibinfo {volume} {93}},\
  \bibinfo {pages} {233903} (\bibinfo {year} {2004})}\BibitemShut {NoStop}%
\bibitem [{\citenamefont {Wang}\ \emph
  {et~al.}(2020{\natexlab{a}})\citenamefont {Wang}, \citenamefont {Zheng},
  \citenamefont {Chen}, \citenamefont {Huang}, \citenamefont {Kartashov},
  \citenamefont {Torner}, \citenamefont {Konotop},\ and\ \citenamefont
  {Ye}}]{nature.577.7788}%
  \BibitemOpen
  \bibfield  {author} {\bibinfo {author} {\bibfnamefont {P.}~\bibnamefont
  {Wang}}, \bibinfo {author} {\bibfnamefont {Y.}~\bibnamefont {Zheng}},
  \bibinfo {author} {\bibfnamefont {X.}~\bibnamefont {Chen}}, \bibinfo {author}
  {\bibfnamefont {C.}~\bibnamefont {Huang}}, \bibinfo {author} {\bibfnamefont
  {Y.~V.}\ \bibnamefont {Kartashov}}, \bibinfo {author} {\bibfnamefont
  {L.}~\bibnamefont {Torner}}, \bibinfo {author} {\bibfnamefont {V.~V.}\
  \bibnamefont {Konotop}},\ and\ \bibinfo {author} {\bibfnamefont
  {F.}~\bibnamefont {Ye}},\ }\bibfield  {title} {\bibinfo {title} {Localization
  and delocalization of light in photonic moir\'{e} lattices},\ }\href
  {https://doi.org/10.1038/s41586-019-1851-6} {\bibfield  {journal} {\bibinfo
  {journal} {Nature (London)}\ }\textbf {\bibinfo {volume} {577}},\ \bibinfo
  {pages} {42} (\bibinfo {year} {2020}{\natexlab{a}})}\BibitemShut {NoStop}%
\bibitem [{\citenamefont {He}\ \emph {et~al.}(2021)\citenamefont {He},
  \citenamefont {Mao}, \citenamefont {Cai}, \citenamefont {Zhang},
  \citenamefont {Li}, \citenamefont {Yuan}, \citenamefont {Zhu},\ and\
  \citenamefont {Wang}}]{PhysRevLett.126.103601}%
  \BibitemOpen
  \bibfield  {author} {\bibinfo {author} {\bibfnamefont {Y.}~\bibnamefont
  {He}}, \bibinfo {author} {\bibfnamefont {R.}~\bibnamefont {Mao}}, \bibinfo
  {author} {\bibfnamefont {H.}~\bibnamefont {Cai}}, \bibinfo {author}
  {\bibfnamefont {J.-X.}\ \bibnamefont {Zhang}}, \bibinfo {author}
  {\bibfnamefont {Y.}~\bibnamefont {Li}}, \bibinfo {author} {\bibfnamefont
  {L.}~\bibnamefont {Yuan}}, \bibinfo {author} {\bibfnamefont {S.-Y.}\
  \bibnamefont {Zhu}},\ and\ \bibinfo {author} {\bibfnamefont {D.-W.}\
  \bibnamefont {Wang}},\ }\bibfield  {title} {\bibinfo {title} {Flat-band
  localization in {C}reutz superradiance lattices},\ }\href
  {https://doi.org/10.1103/PhysRevLett.126.103601} {\bibfield  {journal}
  {\bibinfo  {journal} {Phys. Rev. Lett.}\ }\textbf {\bibinfo {volume} {126}},\
  \bibinfo {pages} {103601} (\bibinfo {year} {2021})}\BibitemShut {NoStop}%
\bibitem [{\citenamefont {Peng}\ \emph {et~al.}(2017)\citenamefont {Peng},
  \citenamefont {Bao},\ and\ \citenamefont {von Oppen}}]{PhysRevB.95.235143}%
  \BibitemOpen
  \bibfield  {author} {\bibinfo {author} {\bibfnamefont {Y.}~\bibnamefont
  {Peng}}, \bibinfo {author} {\bibfnamefont {Y.}~\bibnamefont {Bao}},\ and\
  \bibinfo {author} {\bibfnamefont {F.}~\bibnamefont {von Oppen}},\ }\bibfield
  {title} {\bibinfo {title} {Boundary {Green} functions of topological
  insulators and superconductors},\ }\href
  {https://doi.org/10.1103/PhysRevB.95.235143} {\bibfield  {journal} {\bibinfo
  {journal} {Phys. Rev. B}\ }\textbf {\bibinfo {volume} {95}},\ \bibinfo
  {pages} {235143} (\bibinfo {year} {2017})}\BibitemShut {NoStop}%
\bibitem [{\citenamefont {Martinez~Alvarez}\ and\ \citenamefont
  {Coutinho-Filho}(2019)}]{PhysRevA.99.013833}%
  \BibitemOpen
  \bibfield  {author} {\bibinfo {author} {\bibfnamefont {V.~M.}\ \bibnamefont
  {Martinez~Alvarez}}\ and\ \bibinfo {author} {\bibfnamefont {M.~D.}\
  \bibnamefont {Coutinho-Filho}},\ }\bibfield  {title} {\bibinfo {title} {Edge
  states in trimer lattices},\ }\href
  {https://doi.org/10.1103/PhysRevA.99.013833} {\bibfield  {journal} {\bibinfo
  {journal} {Phys. Rev. A}\ }\textbf {\bibinfo {volume} {99}},\ \bibinfo
  {pages} {013833} (\bibinfo {year} {2019})}\BibitemShut {NoStop}%
\bibitem [{\citenamefont {Andrea~Melloni}\ and\ \citenamefont
  {Martinelli}(2008)}]{OL.33.002389}%
  \BibitemOpen
  \bibfield  {author} {\bibinfo {author} {\bibfnamefont {C.~F.}\ \bibnamefont
  {Andrea~Melloni}, \bibfnamefont {Francesco~Morichetti}}\ and\ \bibinfo
  {author} {\bibfnamefont {M.}~\bibnamefont {Martinelli}},\ }\bibfield  {title}
  {\bibinfo {title} {Continuously tunable 1 byte delay in coupled-resonator
  optical waveguides},\ }\href {https://doi.org/10.1364/OL.33.002389}
  {\bibfield  {journal} {\bibinfo  {journal} {Opt. Lett.}\ }\textbf {\bibinfo
  {volume} {33}},\ \bibinfo {pages} {2389} (\bibinfo {year}
  {2008})}\BibitemShut {NoStop}%
\bibitem [{\citenamefont {Hafezi}\ \emph {et~al.}(2013)\citenamefont {Hafezi},
  \citenamefont {Mittal}, \citenamefont {Fan}, \citenamefont {Migdall},\ and\
  \citenamefont {Taylor}}]{nphoton.2013.274}%
  \BibitemOpen
  \bibfield  {author} {\bibinfo {author} {\bibfnamefont {M.}~\bibnamefont
  {Hafezi}}, \bibinfo {author} {\bibfnamefont {S.}~\bibnamefont {Mittal}},
  \bibinfo {author} {\bibfnamefont {J.}~\bibnamefont {Fan}}, \bibinfo {author}
  {\bibfnamefont {A.}~\bibnamefont {Migdall}},\ and\ \bibinfo {author}
  {\bibfnamefont {J.~M.}\ \bibnamefont {Taylor}},\ }\bibfield  {title}
  {\bibinfo {title} {Imaging topological edge states in silicon photonics},\
  }\href {https://doi.org/10.1038/nphoton.2013.274} {\bibfield  {journal}
  {\bibinfo  {journal} {Nat. Photonics}\ }\textbf {\bibinfo {volume} {7}},\
  \bibinfo {pages} {1001} (\bibinfo {year} {2013})}\BibitemShut {NoStop}%
\bibitem [{\citenamefont {Wang}\ \emph {et~al.}(2015)\citenamefont {Wang},
  \citenamefont {Yao},\ and\ \citenamefont {Poon}}]{fmats.2015.00034}%
  \BibitemOpen
  \bibfield  {author} {\bibinfo {author} {\bibfnamefont {J.}~\bibnamefont
  {Wang}}, \bibinfo {author} {\bibfnamefont {Z.}~\bibnamefont {Yao}},\ and\
  \bibinfo {author} {\bibfnamefont {A.~W.}\ \bibnamefont {Poon}},\ }\bibfield
  {title} {\bibinfo {title} {Silicon-nitride-based integrated optofluidic
  biochemical sensors using a coupled-resonator optical waveguide},\ }\bibfield
   {journal} {\bibinfo  {journal} {Front. Mater.}\ }\textbf {\bibinfo {volume}
  {2}},\ \href {https://doi.org/10.3389/fmats.2015.00034}
  {10.3389/fmats.2015.00034} (\bibinfo {year} {2015})\BibitemShut {NoStop}%
\bibitem [{\citenamefont {Wang}\ \emph
  {et~al.}(2020{\natexlab{b}})\citenamefont {Wang}, \citenamefont {Yao},
  \citenamefont {Wu}, \citenamefont {Fang}, \citenamefont {Lv}, \citenamefont
  {Zhang}, \citenamefont {Lin}, \citenamefont {Fang},\ and\ \citenamefont
  {Cheng}}]{NJP.22.7.2020}%
  \BibitemOpen
  \bibfield  {author} {\bibinfo {author} {\bibfnamefont {M.}~\bibnamefont
  {Wang}}, \bibinfo {author} {\bibfnamefont {N.}~\bibnamefont {Yao}}, \bibinfo
  {author} {\bibfnamefont {R.}~\bibnamefont {Wu}}, \bibinfo {author}
  {\bibfnamefont {Z.}~\bibnamefont {Fang}}, \bibinfo {author} {\bibfnamefont
  {S.}~\bibnamefont {Lv}}, \bibinfo {author} {\bibfnamefont {J.}~\bibnamefont
  {Zhang}}, \bibinfo {author} {\bibfnamefont {J.}~\bibnamefont {Lin}}, \bibinfo
  {author} {\bibfnamefont {W.}~\bibnamefont {Fang}},\ and\ \bibinfo {author}
  {\bibfnamefont {Y.}~\bibnamefont {Cheng}},\ }\bibfield  {title} {\bibinfo
  {title} {Strong nonlinear optics in on-chip coupled lithium niobate microdisk
  photonic molecules},\ }\href {https://doi.org/10.1088/1367-2630/ab97ea}
  {\bibfield  {journal} {\bibinfo  {journal} {New J. Phys.}\ }\textbf {\bibinfo
  {volume} {22}},\ \bibinfo {pages} {073030} (\bibinfo {year}
  {2020}{\natexlab{b}})}\BibitemShut {NoStop}%
\bibitem [{\citenamefont {Kors}\ \emph {et~al.}(2018)\citenamefont {Kors},
  \citenamefont {Reithmaier},\ and\ \citenamefont
  {Benyoucef}}]{apl.112.17.2018}%
  \BibitemOpen
  \bibfield  {author} {\bibinfo {author} {\bibfnamefont {A.}~\bibnamefont
  {Kors}}, \bibinfo {author} {\bibfnamefont {J.~P.}\ \bibnamefont
  {Reithmaier}},\ and\ \bibinfo {author} {\bibfnamefont {M.}~\bibnamefont
  {Benyoucef}},\ }\bibfield  {title} {\bibinfo {title} {Telecom wavelength
  single quantum dots with very small excitonic fine-structure splitting},\
  }\href {https://doi.org/10.1063/1.5023184} {\bibfield  {journal} {\bibinfo
  {journal} {Appl. Phys. Lett.}\ }\textbf {\bibinfo {volume} {112}},\ \bibinfo
  {pages} {172102} (\bibinfo {year} {2018})}\BibitemShut {NoStop}%
\bibitem [{\citenamefont {Haffouz}\ \emph {et~al.}(2018)\citenamefont
  {Haffouz}, \citenamefont {Zeuner}, \citenamefont {Dalacu}, \citenamefont
  {Poole}, \citenamefont {Lapointe}, \citenamefont {Poitras}, \citenamefont
  {Mnaymneh}, \citenamefont {Wu}, \citenamefont {Couillard}, \citenamefont
  {Korkusinski}, \citenamefont {Sch\"{o}ll}, \citenamefont {J\"{o}ns},
  \citenamefont {Zwiller},\ and\ \citenamefont
  {Williams}}]{nanolett.18.5.2018}%
  \BibitemOpen
  \bibfield  {author} {\bibinfo {author} {\bibfnamefont {S.}~\bibnamefont
  {Haffouz}}, \bibinfo {author} {\bibfnamefont {K.~D.}\ \bibnamefont {Zeuner}},
  \bibinfo {author} {\bibfnamefont {D.}~\bibnamefont {Dalacu}}, \bibinfo
  {author} {\bibfnamefont {P.~J.}\ \bibnamefont {Poole}}, \bibinfo {author}
  {\bibfnamefont {J.}~\bibnamefont {Lapointe}}, \bibinfo {author}
  {\bibfnamefont {D.}~\bibnamefont {Poitras}}, \bibinfo {author} {\bibfnamefont
  {K.}~\bibnamefont {Mnaymneh}}, \bibinfo {author} {\bibfnamefont
  {X.}~\bibnamefont {Wu}}, \bibinfo {author} {\bibfnamefont {M.}~\bibnamefont
  {Couillard}}, \bibinfo {author} {\bibfnamefont {M.}~\bibnamefont
  {Korkusinski}}, \bibinfo {author} {\bibfnamefont {E.}~\bibnamefont
  {Sch\"{o}ll}}, \bibinfo {author} {\bibfnamefont {K.~D.}\ \bibnamefont
  {J\"{o}ns}}, \bibinfo {author} {\bibfnamefont {V.}~\bibnamefont {Zwiller}},\
  and\ \bibinfo {author} {\bibfnamefont {R.~L.}\ \bibnamefont {Williams}},\
  }\bibfield  {title} {\bibinfo {title} {Bright single inasp quantum dots at
  telecom wavelengths in position-controlled inp nanowires: The role of the
  photonic waveguide},\ }\href {https://doi.org/10.1021/acs.nanolett.8b00550}
  {\bibfield  {journal} {\bibinfo  {journal} {Nano Lett.}\ }\textbf {\bibinfo
  {volume} {18}},\ \bibinfo {pages} {3047} (\bibinfo {year}
  {2018})}\BibitemShut {NoStop}%
\bibitem [{\citenamefont {Arakawa}\ and\ \citenamefont
  {Holmes}(2020)}]{APR.7.2.2020}%
  \BibitemOpen
  \bibfield  {author} {\bibinfo {author} {\bibfnamefont {Y.}~\bibnamefont
  {Arakawa}}\ and\ \bibinfo {author} {\bibfnamefont {M.~J.}\ \bibnamefont
  {Holmes}},\ }\bibfield  {title} {\bibinfo {title} {Progress in quantum-dot
  single photon sources for quantum information technologies: A broad spectrum
  overview},\ }\href {https://doi.org/10.1063/5.0010193} {\bibfield  {journal}
  {\bibinfo  {journal} {Appl. Phys. Rev.}\ }\textbf {\bibinfo {volume} {7}},\
  \bibinfo {pages} {021309} (\bibinfo {year} {2020})}\BibitemShut {NoStop}%
\bibitem [{\citenamefont {Htoon}\ \emph {et~al.}(2002)\citenamefont {Htoon},
  \citenamefont {Takagahara}, \citenamefont {Kulik}, \citenamefont {Baklenov},
  \citenamefont {Holmes},\ and\ \citenamefont {Shih}}]{PhysRevLett.88.087401}%
  \BibitemOpen
  \bibfield  {author} {\bibinfo {author} {\bibfnamefont {H.}~\bibnamefont
  {Htoon}}, \bibinfo {author} {\bibfnamefont {T.}~\bibnamefont {Takagahara}},
  \bibinfo {author} {\bibfnamefont {D.}~\bibnamefont {Kulik}}, \bibinfo
  {author} {\bibfnamefont {O.}~\bibnamefont {Baklenov}}, \bibinfo {author}
  {\bibfnamefont {A.~L.}\ \bibnamefont {Holmes}},\ and\ \bibinfo {author}
  {\bibfnamefont {C.~K.}\ \bibnamefont {Shih}},\ }\bibfield  {title} {\bibinfo
  {title} {Interplay of rabi oscillations and quantum interference in
  semiconductor quantum dots},\ }\href
  {https://doi.org/10.1103/PhysRevLett.88.087401} {\bibfield  {journal}
  {\bibinfo  {journal} {Phys. Rev. Lett.}\ }\textbf {\bibinfo {volume} {88}},\
  \bibinfo {pages} {087401} (\bibinfo {year} {2002})}\BibitemShut {NoStop}%
\bibitem [{\citenamefont {Xu}\ \emph {et~al.}(2008)\citenamefont {Xu},
  \citenamefont {Fattal},\ and\ \citenamefont {Beausoleil}}]{Xu:08}%
  \BibitemOpen
  \bibfield  {author} {\bibinfo {author} {\bibfnamefont {Q.}~\bibnamefont
  {Xu}}, \bibinfo {author} {\bibfnamefont {D.}~\bibnamefont {Fattal}},\ and\
  \bibinfo {author} {\bibfnamefont {R.~G.}\ \bibnamefont {Beausoleil}},\
  }\bibfield  {title} {\bibinfo {title} {Silicon microring resonators with
  1.5-{\textmu}m radius},\ }\href {https://doi.org/10.1364/OE.16.004309}
  {\bibfield  {journal} {\bibinfo  {journal} {Opt. Express}\ }\textbf {\bibinfo
  {volume} {16}},\ \bibinfo {pages} {4309} (\bibinfo {year}
  {2008})}\BibitemShut {NoStop}%
\end{thebibliography}

\providecommand{\noopsort}[1]{}\providecommand{\singleletter}[1]{#1}%

\end{document}